\documentclass[10pt,runningheads]{llncs}
\usepackage[T1]{fontenc}
\usepackage{graphicx}
\usepackage{hyperref}       
\usepackage{url}            
\usepackage{booktabs}       
\usepackage{amsfonts}       
\usepackage{nicefrac}       
\usepackage{microtype}      
\usepackage{xcolor}         

\usepackage{graphicx}%
\usepackage{multirow}%
\usepackage{amsmath,amssymb,amsfonts}%
\usepackage{mathrsfs}%
\usepackage{algorithm}%
\usepackage{algorithmicx}%
\usepackage{algpseudocode}%
\usepackage{listings}%
\usepackage{mathtools}
\usepackage{array}
\usepackage{color}

\usepackage[margin=1in]{geometry}

\newcommand{\cD}{\mathcal{D}}
\newcommand{\cC}{\mathcal{C}}
\newcommand{\cA}{\mathcal{A}}

\newcommand{\cN}{\mathcal{N}}
\newcommand{\cG}{\mathcal{G}}
\newcommand{\cL}{\mathcal{L}}
\newcommand{\cO}{\ensuremath{\mathcal{O}}}
\newcommand{\cX}{\ensuremath{\mathcal{X}}}
\newcommand{\cY}{\ensuremath{\mathcal{Y}}}

\newcommand{\cP}{\ensuremath{\mathcal{P}}}

\newcommand{\Mij}[2]{\ensuremath{M_{{#1} \rightarrow {#2}}}}
\newcommand{\bR}{\mathbb{R}}

\newcommand{\tr}{\mathsf{tr}}

\newcommand{\rr}{\mathsf{RR}}

\newcommand{\pr}[1]{\mathbb{P}\left[ #1 \right]}
\newcommand{\ex}[2]{\underset{#1}{\mathbb{E}}\left[{#2}\right]}

\usepackage{refcount}

\newif\ifnewtext
\newtextfalse
\ifnewtext
    \newenvironment{modtext}
    {\color{blue}} 
    {} 
\else
    \newenvironment{modtext}
    {} 
    {} 
\fi
\begin{document}
\title{GOPHER: Optimization-based Phenotype Randomization for Genome-Wide Association Studies with Differential Privacy}
\titlerunning{Phenotype Randomization for GWAS with DP}
%
\author{Anupama Nandi\inst{1} \and
Seth Neel\inst{2} \and
Hyunghoon Cho\inst{1,3}}
%
%
\institute{Department of Biomedical Informatics \& Data Science, Yale School of Medicine, New Haven, CT, USA 
\and
Department of Technology and Operations Management, Harvard Business School, Boston, MA, USA \\
\and
Department of Computer Science, Yale University, New Haven, CT, USA
}
\maketitle              
\begin{abstract}
Genome-wide association studies (GWAS) are an essential tool in biomedical research for identifying genetic factors linked to health and disease. However, publicly releasing GWAS summary statistics poses well-recognized privacy risks, including the potential to infer an individual's participation in the study or to reveal sensitive phenotypic information (e.g., disease status). While differential privacy (DP) offers a rigorous mathematical framework for mitigating these risks, existing DP techniques for GWAS either introduce excessive noise or restrict the release to a limited set of results. In this work, we present practical DP mechanisms for releasing the complete set of genome-wide association statistics with privacy guarantees. We demonstrate the accuracy of the privacy-preserving statistics released by our mechanisms on a range of GWAS datasets from the UK Biobank, utilizing both real and simulated phenotypes. We introduce two key techniques to overcome the limitations of prior approaches: (1) an optimization-based randomization mechanism that directly minimizes the expected error in GWAS results to enhance utility, and (2) the use of personalized priors, derived from predictive models privately trained on a subset of the dataset, to enable sample-specific optimization which further reduces the amount of noise introduced by DP. Overall, our work provides practical tools for accurately releasing comprehensive GWAS results with provable protection of study participants.

\end{abstract}

\newpage
\section{Introduction}
\label{sec:intro}
Genome-wide association studies (GWAS) employ statistical tests over groups of individuals to identify genetic variants associated with traits of interest. However, the use of individual-level genomic data and associated sensitive personal information, such as health status and demographic attributes, in these studies raises important privacy concerns. It is well-established that even when only aggregate-level association statistics are released, these data can be exploited to infer sensitive information about individuals in the study.
For example, given access to a target individual's genotypes, an adversary can exploit the released statistics to infer whether that individual was part of the study cohort---known as a \emph{membership inference attack}---which may, in turn, reveal sensitive health information~\cite{Homer,Wang2009Learning}. In response to these risks, the National Institutes of Health (NIH) has established genomic data sharing policies that restrict access to aggregate statistics from studies involving especially sensitive cohorts, such as those associated with stigmatized conditions or vulnerable populations \cite{PotentialRA}. Such restrictions to data sharing introduce significant barriers for researchers seeking to access and utilize these valuable statistical data.

Differential privacy (DP)~\cite{DMNS06,DKMMN06} is a widely adopted framework for privacy protection across diverse domains, offering a principled approach to safeguarding individual-level information in GWAS. However, the extremely high dimensionality of GWAS results, often comprising hundreds of thousands to millions of statistics (corresponding to genetic variants), poses significant challenges for applying DP effectively, which typically introduces noise that scales with the amount of data released. While several approaches have been developed for sharing genomic summary statistics with DP~\cite{johnson2013privacy,yu2014scalable,han_logistic,yu2014differentially,Simmons2018ProtectingGD}, these methods require adding substantial noise to achieve reasonable privacy, resulting in severely degraded accuracy. Due to these challenges, prior studies focused on restricted scenarios where only a small set (e.g., 5-10) of the strongest associations are released. However, this greatly limits the utility of GWAS results, especially given that many downstream analyses, such as disease risk prediction and meta-analyses, depend on access to the full genome-wide statistics.

In this work, we present GOPHER (GWAS with Optimization-based PHEnotype Randomization), a set of utility-maximizing DP mechanisms capable of releasing the complete set of GWAS statistics with enhanced accuracy while providing DP guarantees. Following prior work~\cite{PrivSTRAT}, we adopt the notion of phenotypic DP (analogous to label DP~\cite{labeldp,badanidiyuru2023,esfandiari2022label} in the privacy literature), which treats phenotypes as sensitive attributes to protect, while excluding genotypes. This choice reflects the common adversarial model where the attacker  has access to a target individual's genotypes and attempts to infer additional attributes (e.g., cohort membership or health status) based on the released results.
Full DP over both genotypes and phenotypes is ideal but remains impractical; phenotypic DP offers a practical middle ground, enabling accurate, full GWAS releases  while protecting individuals under the primary adversarial model of concern.

Our approach focuses on randomizing individual phenotypes to achieve DP. Once randomized, the GWAS analysis proceeds as in the standard (non-private) setting, producing statistics that inherit DP guarantees via the post-processing property of DP, which states that any computation on DP-protected data that does not involve additional private inputs remains differentially private \cite{Dwork2014}. Conceptually, our method resembles the randomized response (RR) mechanism~\cite{ghazi2023regression}, as it randomizes each phenotype value according to some function. However, we apply this strategy at the dataset level by jointly optimizing the randomization functions to leverage the non-private information about the genotype matrix.

We introduce two key strategies to enhance utility. First, we model the entire GWAS workflow and formulate an optimization problem to construct randomization mechanisms---i.e., conditional distributions for sampling perturbed phenotypes for each individual---that minimize the expected error in the final statistics. This allows us to surpass baseline methods, such as RR, which do not account for the downstream use of noisy outputs in GWAS. Second, as a central component of our optimization, we incorporate \emph{personalized} priors on individual phenotypes, derived from predictive models (e.g., polygenic risk scores~\cite{PRScs}) trained on genotypes to estimate the phenotype. This enables the DP mechanism to be tailored to each individual. Our approach notably departs from existing work on optimal mechanisms for label DP \cite{ghazi2023regression}, where a single global mechanism is applied uniformly across all samples. Our strategy of adding sample-specific noise by leveraging predictive modeling may be of broad interest beyond GWAS.

Across a broad evaluation spanning real and simulated phenotypes for 100,000 individuals from the UK Biobank, we demonstrate that our mechanisms yield substantial utility gains. The resulting association statistics accurately match those from non-private GWAS while maintaining rigorous privacy guarantees. Notably, our work substantially advances the scalability of differentially private GWAS, enabling the release of full summary statistics (e.g., over 500,000 variants) instead of the small number of top associations permitted by previous methods. In addition, our phenotype-randomization framework supports an arbitrary number of downstream analyses using the privatized phenotypes with no additional privacy cost, further expanding the practical utility of our approach.
Our open-source software is available at: \url{https://github.com/hcholab/gopher}.

\section{Methods}
\label{sec:methods}

\subsection{Review of Association Testing in GWAS} \label{subsec:gwas_and_dp}
Consider a dataset of $n$ individuals genotyped at $m$ variants. 
Let $\mathbf{X} =[x_{ij}]$ be the $n \times m$ sample-by-variant genotype matrix, where $x_{ij} \in \{0,1,2\}$ is the minor allele count for individual $i$ at the $j$-th variant. Next, let $\mathbf{y} = (y_i)_{i=1}^{n}$ be the phenotype vector of length $n$, where $y_i$ is the phenotypic value of individual $i$. 
We assess the association between each
variant and phenotype of interest based on the following standard linear regression model for GWAS:
\begin{align*}
    \mathbf{y}  =  \mathbf{\Tilde{x}}_{*,k}\beta_k + e,
\end{align*}
where $\mathbf{\Tilde{x}}_{*,k}$ denotes the column of $\mathbf{X}$ at the $k$-th variant after standardization,  $\beta_k$ denotes fixed effect for variant $k$, and  $e \sim \cN(0,\sigma_e^2 \mathbf{I}_{n\times n})$ accounts for residual errors representing environmental noise.
The Pearson correlation coefficient ($r$) between the genotype and phenotype vectors is utilized to quantify the association between each
SNP and the phenotype of interest, defined as follows:
\begin{align*}
r_k = \frac{\mathbf{\tilde{x}}_{*,k}^T(\mathbf{y-\mu_{\mathbf{y}}})}{n\sigma_{\mathbf{y}}},
\end{align*}
for the $k$-th variant. This quantity can be converted to a $t$-statistic using the formula $t^2 = r^2 (n - 2) / (1 - r^2)$, which, under the null model (no association), follows a $\chi^2$ distribution with one degree of freedom. From this, $p$-values can be calculated to assess the statistical significance of the association.

\subsection{Review of Differential Privacy}


Let $\boldsymbol{\cX} \times \cY$ be the domain of GWAS data from the previous section. In this work, we focus on samples of the form $(\mathbf{x},y) \in \boldsymbol{\cX} \times \cY$, where $\mathbf{x}$ contains $m$ categorical features (genotypes), i.e., $\boldsymbol{\cX} = \cX_1 \times \cdots \times \cX_m$, and $y$ is a discrete label (phenotype). 
A dataset, denoted by $S$, consists of a multiset of $n$ samples from $\boldsymbol{\cX} \times \cY$.  
We use the following definition of differential privacy:
\begin{definition}[Differential Privacy \cite{DMNS06,DKMMN06}]\label{def:DP}
	Let $\epsilon,\delta >0$. A (randomized) algorithm $\cA:(\boldsymbol{\cX} \times \cY)^n \rightarrow \bR$ is $(\epsilon,\delta)$-differentially private if for any two neighboring datasets $S,S'\in (\boldsymbol{\cX} \times \cY)^n$, which differ only by a single data sample, and for any measurable $\cO \subseteq \bR$, it holds that: 
    \begin{align*}
        \pr{ \cA(S) \in \cO } \leq e^\epsilon \cdot \pr{\cA(S') \in \cO}+\delta.
    \end{align*}
If $\delta=0$, then $\cA$ is said to $\epsilon$-differentially private ($\epsilon$-DP). 
\end{definition} 
In this work, we focus on (pure) $\epsilon$-DP with $\delta = 0$, while adopting a slightly relaxed privacy notion known as \emph{label differential privacy}~\cite{labeldp}, where the neighboring relation is defined only over differences in $y$, rather than jointly over $(\mathbf{x}, y)$.

\begin{definition}[Label Differential Privacy \cite{labeldp}]\label{def:labelDP}
	Let $\epsilon,\delta >0$. A (randomized) algorithm $\cA:(\boldsymbol{\cX} \times \cY)^n \rightarrow \bR$ is $(\epsilon,\delta)$-Label DP if for all pairs of neighboring datasets $S,S'$ differing only in the label of a single data sample, and every measurable $\cO \subseteq \bR$, we have: \begin{align*}
	    \pr{ \cA(S) \in \cO } \leq e^\epsilon \cdot \pr{ \cA(S') \in \cO  +\delta}.
	\end{align*}
\end{definition} 

\noindent Next, we recall some basic DP mechanisms that we use throughout our work.


\begin{lemma}[Laplace Mechanism]
Let $f: (\boldsymbol{\cX} \times \cY)^n \rightarrow \bR^d$ be a function with $\ell_1$-sensitivity $\Delta_{f,1}:= \max_{S\sim S' \in (\boldsymbol{\cX} \times \cY)^n}||f(S)-f(S')||_1$. Then the Laplace mechanism $\cA_f(S) = f(S) + \mathrm{Lap}\left(\frac{\Delta_{f,1}}{\epsilon}\right)$ satisfies $\epsilon$-DP.    
\end{lemma}

\begin{lemma}[Randomized Response (RR) \cite{Warner1965RandomizedRA}]
For any positive integer $k$, let $[k]:={1,\ldots,k}.$ For any $\epsilon \geq 0,$ the randomized response mechanism, denoted by $\rr_{\epsilon,k}$, takes as input $ y \in [k]$, and outputs a random sample $\hat{y}$ drawn from the following probability distribution:
\begin{align*}
 \pr{y =\hat{y}} =
    \begin{cases}
       \frac{e^\epsilon}{e^\epsilon+k-1} , & \text{for } \hat{y} = y\\
        \frac{1}{e^\epsilon+k-1},              & \text{otherwise.}
    \end{cases}
\end{align*}
The output of $\rr_{\epsilon,k}$ satisfies $\epsilon$-DP.
\end{lemma}



\subsection{Prior Approaches to GWAS with Differential Privacy and Their Limitations}
Prior studies have explored differentially private release of key genomic summary statistics, including minor allele frequencies (MAFs), chi-square statistics, association $p$-values, top-$k$ variants, and pairwise variant correlations~\cite{johnson2013privacy,yu2014scalable,uhlerop2013privacy,yu2014iDASH}. These methods typically adapt DP to case-control GWAS using neighbor distance-based, Laplace, or score-based mechanisms, while differentially private logistic regression has also been applied to both variant-disease and variant-variant associations~\cite{han_logistic,yu2014differentially}. 

PrivSTRAT~\cite{PrivSTRAT} addressed population stratification correction under DP and introduced the notion of  \emph{phenotypic differential privacy} to protect sensitive phenotypes, aligning with label DP (Definition~\ref{def:labelDP}), which we also adopt. PrivSTRAT employs the standard Laplace mechanism to compute the association statistic between a top SNP and the phenotype vector. Notably, five widely cited publications on DP-GWAS~\cite{johnson2013privacy,uhlerop2013privacy,simmons2016realizing,tramer2015differential,aziz}, including a recent 2021 study~\cite{aziz}, all employ the Laplace mechanism to independently release the statistic at each genetic variant.  However, these methods focus on releasing only a limited subset of results (e.g., the top 3 or 5 most significantly associated variants) and are not designed to support the release of the complete set of GWAS statistics. Given the large number of variants typically included in GWAS results, existing DP methods would require adding excessive noise, rendering the outputs unusable. In contrast, our phenotype randomization approach decouples noise from the number of variants, addressing a significant scalability challenge that, to our knowledge, has not been previously explored in the literature. 

We discuss additional related work on alternative privacy notions and general label DP techniques in Supplementary Note~\hyperref[appdx:priv-strat]{S1}.


\subsection{Our Optimization-Based Phenotype Randomization (GOPHER) Framework}
\label{subsec:method}
We present our GOPHER mechanisms for the differentially private release of GWAS statistics. 
In our setting, the phenotype (label) vector $\mathbf{y} \in \mathcal{Y}^{n \times 1} \subset \mathbb{R}^{n}$ is treated as private, while the genotype (feature) matrix $\mathbf{X} \subset \mathbb{R}^{n \times m}$ is assumed public. Consequently, the output statistics are required to be differentially private with respect to $\mathbf{y}$ only. This corresponds to the notion of \emph{phenotypic differential privacy}, introduced in prior work~\cite{PrivSTRAT}.
In the following, our discussion of DP will be confined to the concept of phenotypic DP. 



\vspace{0.25em}
\begin{definition}[Phenotypic Differential Privacy \cite{PrivSTRAT}]
Let $\cA$ be a randomized algorithm taking in as input a $n \times m$ genotype matrix, $\mathbf{X}$, and an $n$-dimensional phenotype vector, $\mathbf{y}$. We say that $\cA$ is $\epsilon$-phenotypic differentially private if it satisfies $\epsilon$-DP with respect to all neighboring datasets $(\mathbf{X}, \mathbf{y})$ and $(\mathbf{X}, \mathbf{y}')$ for all genotype matrices $\mathbf{X}$ such that $\mathbf{y}$ and $\mathbf{y}'$ differ in exactly one individual.
\end{definition}

Our approach introduces optimization-based mechanisms that perturb each label individually using a DP randomization function. The privatized labels are then used to compute the desired statistics via post-processing, ensuring that the overall mechanism satisfies DP with respect to $\mathbf{y}$. Our work is inspired by the general utility maximization framework in DP~\cite{ghazi2023regression,ghosh2012}, which involves two key steps: (i) parameterizing the space of DP mechanisms, and (ii) optimizing these parameters to maximize utility with respect to a given loss function. For regression objectives such as squared or Poisson-log loss, prior work~\cite{ghazi2023regression} proposed a label-DP randomizer that minimizes the expected loss between true and noisy labels under a prior distribution over the true labels, without considering input features. They showed that the optimal randomizer $\mathcal{A}$ takes the form of a randomized response mechanism over discretized (binned) label values (see Supplementary Note~\hyperref[appdx:rr-on-bins]{S2} for details).

Our first algorithm, GOPHER-LP (Algorithm~\ref{alg:gopher-lp}), adapts this framework to the GWAS setting under phenotypic DP. We first estimate a prior distribution over phenotype values using the private histogram method~\cite{ghazi2023regression}. Using this private prior, the observed phenotype data, and the expected squared Euclidean distance between the true and randomized phenotypes as the utility measure, we formulate a linear programming (LP) optimization problem to derive the optimal DP randomization mechanism. The perturbed phenotypes sampled from this mechanism are then used for GWAS.



\begin{algorithm}[tb]
\small
    \caption{\textbf{GOPHER-LP}}
    \label{alg:gopher-lp}
\begin{algorithmic}[1]
    \State \textbf{Input:} Genotypes $\mathbf{X} \in \mathbb{R}^{n \times m}$, Phenotypes $\mathbf{y} \in \mathcal{Y}^n \subset \mathbb{R}^{n}$, Privacy parameters $\epsilon_1, \epsilon_2$, Number of bins $b$
    
    \State \textbf{Output:} Randomized phenotypes $\hat{y}_1, \ldots, \hat{y}_n \in \hat{\mathcal{Y}}$ satisfying $(\epsilon_1+\epsilon_2)$-phenotypic DP
    \State Let $L \leftarrow \min(\mathcal{Y})$ and $U \leftarrow \max(\mathcal{Y})$.   
    \State Compute $\Delta = (U-L)/(b-1)$.
    \State Define $\cY' = \hat{\cY} := (L,L+\Delta,\ldots,U-\Delta,U).$
    \State Estimate the prior over $\cY'$ via the private histogram mechanism: $\mathbb{P} \leftarrow \cA_{\text{Hist}}(y_1,\ldots y_n;  \epsilon_1)$.
    \State Solve the following LP optimization over the parameters $M:=\{\Mij{u}{v}\}_{u \in \cY',v \in \hat{\cY}}:$
    {\setlength{\abovedisplayskip}{2pt}
     \setlength{\belowdisplayskip}{2pt}
     \small \begin{equation} 
    \begin{aligned} \nonumber
    \text{minimize}_{M} \quad &  \textstyle \sum_{u \in \cY'}\pr{u} \left(\sum_{v \in \hat{\cY}} \Mij{u}{v}\cdot(u-v)^2\right), \\ 
    \textrm{subject to} \quad & \forall  u  \in \cY',v \in \hat{\cY}: \quad\quad  \Mij{u}{v} \geq 0,  \\
      & \forall u \in \cY': \quad  \textstyle \sum_{v \in \hat{\cY}} \Mij{u}{v} = 1,   \\ 
      & \forall v \in \hat{\cY}, \forall u,u' \in \cY', u \neq u', : \quad \Mij{u'}{v} \leq e^{\epsilon_2} \Mij{u}{v}. 
    \end{aligned}
    \end{equation}}
    \For{$i \in [n]$}
    \State Sample $\hat{y}_i$ with probability $\pr{\hat{y}_i = v} = \Mij{u'}{v}$ where $u'$ is the binned value of $y_i$ in $\cY'$.
    \EndFor
\end{algorithmic}
\end{algorithm}


\vspace{-0.5em}
\subsection{GOPHER-MultiLP: Incorporating Personalized Priors}\label{subsec:multilp}
\vspace{-0.5em}
In GOPHER-LP, the prior distribution over phenotypes is estimated without incorporating the public genotype matrix $X$, resulting in a global randomization function applied uniformly to all individuals. While this yields an optimal mechanism within the class of global randomizers, it is likely suboptimal compared to more flexible mechanisms that leverage available non-private information to tailor randomization at the individual level. To address this, we propose GOPHER-MultiLP, which incorporates individual-specific information by perturbing each $y_i$ using an LP problem informed by a personalized prior $\mathbb{P}[y_i | \mathbf{x}_i]$, conditioned on the individual's genotype.

Assuming access to an oracle for the joint conditional distribution $\mathbb{P}[\mathbf{y} | \mathbf{X}]$ and the simplifying assumption that phenotypes are conditionally independent given genotypes, we approximate the marginal prior for each individual as $\mathbb{P}[y_i | \mathbf{x}_i]$. Using this personalized prior, we apply GOPHER-LP independently to each data sample $(\mathbf{x}_i, y_i)$, solving the corresponding LP to obtain an adaptive randomization mechanism that incorporates genotype information during optimization. This leads to improved utility compared to the genotype-agnostic GOPHER-LP, as we show in our experiments.

In practice, we approximate the oracle using a held-out subset of the data. We first select a small subset $(\mathbf{X}_1, \mathbf{y}_1)$ and privately estimate the initial prior over $\mathbf{y}_1$ via the private histogram method, as described earlier. Applying GOPHER-LP, we generate privatized phenotypes $\mathbf{\hat{y}}_1$, which are then used to estimate the conditional prior. Assuming a linear phenotype model $y_i = \sum_{j} x_{i,j}\beta_j + e_i$, we estimate the fixed-effect coefficients $\boldsymbol{\beta}$ and the residual variance $\sigma_y^2$, resulting in $\mathbb{P}[y_i | \mathbf{x}_i] = \mathcal{N}(\boldsymbol{\beta}^\top \mathbf{x}_i, \sigma_y^2)$. Any statistical method can be used to estimate $\boldsymbol{\beta}$, including recent polygenic risk score (PRS) methods such as PRS-CS~\cite{PRScs}, which we use in our experiments.

Finally, we use the updated, individual-specific prior $\mathbb{P}[y_i |\mathbf{x}_i]$ to generate optimally perturbed phenotypes for the remaining dataset. Since each sample is used only once within our held-out workflow, the overall privacy cost remains unchanged due to the parallel composition property of DP~\cite{privacybook}. Note that, we use small portion of the privacy budget to estimate DP variance parameter required for the prior. The pseudocode for this adaptive mechanism, GOPHER-MultiLP, is provided in Algorithm~\ref{alg:gopher-multilp}.

\begin{modtext}
Phenotype prediction is inherently noisy, and genotypes explain only a portion of phenotypic variance. To mitigate the impact of noisy predictions, rather than relying solely on PRS estimates, we extend our approach by grouping individuals according to their inferred phenotypes. We then privately estimate the phenotype mean and variance within each group using a standard Laplace mechanism, constructing a prior that captures broad genotype-phenotype trends while remaining robust to PRS noise. This process yields a personalized prior for each individual reflecting the phenotype distribution of their predicted risk group.
\end{modtext}

\begin{algorithm}[tb]
\small
    \caption{\textbf{GOPHER-MultiLP}}
    \label{alg:gopher-multilp}
\begin{algorithmic}[1]
    \State \textbf{Input:} Genotypes $\mathbf{X} \in \mathbb{R}^{n \times m}$, Phenotypes $\mathbf{y} \in \mathcal{Y}^n \subset \mathbb{R}^{n}$, Privacy parameters $\epsilon_1, \epsilon_2$, Number of bins $b$
    
    \State \textbf{Output:} Randomized phenotypes $\hat{y}_1, \ldots, \hat{y}_n \in \hat{\mathcal{Y}}$ satisfying $(\epsilon_1+\epsilon_2)$-phenotypic DP
  
    \State Let $L \leftarrow \min(\cY)$ , $U \leftarrow \max(\cY)$.   
    \State Compute $\Delta = (U-L)/(b-1)$.
    \State Define $\cY' = \hat{\cY} := (L,L+\Delta,\ldots,U-\Delta,U).$
    \State Randomly split $(\mathbf{X},\bf{y})$ into two subsets $(\mathbf{X}_1,\mathbf{y}_1)$ and $(\mathbf{X}_2,\mathbf{y}_2)$.
    \State $\mathbf{\hat{y}}_1 \leftarrow \text{GOPHER-LP}(\mathbf{X}_1,\mathbf{y}_1, \epsilon_1,\epsilon_2)$.
    \State Train a phenotype prediction (PRS) model: $P_\theta(y|\mathbf{x}) \leftarrow \cA_{\text{PRS}}(\mathbf{X}_1,\mathbf{\hat{y}}_1)$.
    \State Compute personalized priors over $\mathbf{y}_2$: $P_i := P_\theta(y_i|\mathbf{x}_i),\ \forall y_i\in \mathbf{y}_2$.
    \For{$y_i \in \mathbf{y}_2$}
    \State Solve the LP in GOPHER-LP (Algorithm~\ref{alg:gopher-lp}) with parameters $\{\Mij{u}{v}^{(i)} \}_{u \in \cY',v \in \hat{\cY}}$, 
    prior $P_i$, and $\epsilon = \epsilon_1+ \epsilon_2$.
    \State Sample $\hat{y}_i$ with probability $\pr{\hat{y}_i = v} = \Mij{u'}{v}^{(i)}$ where $u'$ is the binned value of $y_i$ in $\cY'$.
    \EndFor
\end{algorithmic}
\end{algorithm}

\subsection{GOPHER-MultiQP: Maximizing Accuracy of Association Statistics}\label{subsec:multiqp}
GOPHER-LP and GOPHER-MultiLP leverage the insight that regression models with label DP will be most effective when minimizing the error introduced in the noisy labels~\cite{ghazi2023regression}. 
However, recall that our goal in GWAS is to estimate the Pearson correlation coefficient, 
 $r_k(\mathbf{x}_{*,k},\mathbf{y}) = \mathbf{\Tilde{x}}_{*,k}^T(\mathbf{y-\mu_{\mathbf{y}}}) / (n \sigma_{\mathbf{y}})$, where $\mathbf{\Tilde{x}}_{*,k}$ (and analogously the full matrix $\Tilde{\mathbf{X}}$) denotes standardized genotypes, which can be precomputed.
Hence, we propose a mechanism to directly optimize the accuracy of these target statistics.

\begin{algorithm}[h]
\small
    \caption{\textbf{GOPHER-MultiQP}}
    \label{alg:gopher-multiqp}
\begin{algorithmic}[1]
        \State \textbf{Input:} Genotypes $\mathbf{X} \in \mathbb{R}^{n \times m}$, Phenotypes $\mathbf{y} \in \mathcal{Y}^n \subset \mathbb{R}^{n}$, Privacy parameters $\epsilon_1, \epsilon_2$, Number of bins $b$, Number of clusters $k$
    
    \State \textbf{Output:} Randomized phenotypes $\hat{y}_1, \ldots, \hat{y}_n \in \hat{\mathcal{Y}}$ satisfying $(\epsilon_1+\epsilon_2)$-phenotypic DP

    \State Let $L \leftarrow \min(\cY)$ , $U \leftarrow \max(\cY)$.   
    \State Compute $\Delta = (U-L)/(b-1)$.
    \State Define $\cY' = \hat{\cY} = (L,L+\Delta,\ldots,U-\Delta,U).$
    \State Randomly split $(\mathbf{X},\bf{y})$ into two subsets $(\mathbf{X}_1,\mathbf{y}_1)$ and $(\mathbf{X}_2,\mathbf{y}_2)$.
    \State $\mathbf{\hat{y}}_1 \leftarrow \text{GOPHER-LP}(\mathbf{X}_1,\mathbf{y}_1, \epsilon_1,\epsilon_2)$.
    \State Train a phenotype prediction (PRS) model: $P_\theta(y|\mathbf{x}) \leftarrow \cA_{\text{PRS}}(\mathbf{X}_1,\mathbf{\hat{y}}_1)$.
    \State Compute personalized priors over $\mathbf{y}_2$: $P_i := P_\theta(y_i|\mathbf{x}_i),\ \forall y_i\in \mathbf{y}_2$.
    \State Apply $k$-means clustering on $\{P_i\}$ to partition $(\mathbf{X}_2, \mathbf{y}_2)$ into $k$ clusters.
    \For{$j = 1$ to $k$}
    \State Let $(\mathbf{X}_2^{(j)}, \mathbf{y}_2^{(j)})$ be the data samples assigned to cluster $j$.
    \State Compute $\mathbf{K}_j=\frac{1}{m}\Tilde{\mathbf{X}}_2^{(j)}\Tilde{\mathbf{X}}_2^{(j)^T}$ using standardized genotypes  $\Tilde{\mathbf{X}}_2^{(j)}$.
    \EndFor
    \begin{modtext}
    \State Compute the block matrix $\mathbf{\Tilde{A}}$ and vector $\mathbf{\Tilde{b}}$ (Supplementary Note~\hyperref[{appdx:phenoQP}]{S3}).
    \State Solve the QP in Eq.~\ref{eqn:multiqp} via the Difference of Convex Algorithm~(Eq.~\ref{eqn:qp_dca}) with parameters $\{\Mij{u}{v}^{(j)} \}_{u \in \cY',v \in \hat{\cY}}$, 
     joint personalized prior $\mathbb{P}[\mathbf{y}_2^{(j)}| \mathbf{X}_2^{(j)}]$ based on $\{P_i\}$ and covariance  matrix $\mathbf{K}_j$, 
     and privacy budget $\epsilon = \epsilon_1+ \epsilon_2$, $\forall j \in [k]$. 
    \For{each cluster $j = 1$ to $k$}
    \For{each $y_i \in \mathbf{y}_2^{(j)}$}
        \State Sample $\hat{y}_i$ with probability $\pr{\hat{y}_i = v} = \Mij{u'}{v}^{(j)}$ where $u'$ is the binned value of $y_i$ in $\cY'$.
    \EndFor
\EndFor
    \end{modtext}
\end{algorithmic}
\end{algorithm}


Let $f({\mathbf{X}},\mathbf{y}) = \Tilde{\mathbf{X}}^T(\mathbf{y-\mu_{\mathbf{y}}}) / (n \sigma_{\mathbf{y}})$ be the function we aim to evaluate and release, encompassing the complete set of correlation coefficients across all variants. 
Let $\cD$ represent the data-generating distribution over $\boldsymbol{\cX} \times \cY$, $\cD_y$ denote the marginal distribution over $y$, and let $\hat{\mathbf{y}}$ signify the output of any $\epsilon$-DP mechanism $\cA$ applied to $\mathbf{y}$. Let $\cY, \hat{\cY} \subseteq \mathbb{R}$ be the input and output sets. 

Our goal is to minimize the expected loss over the final GWAS results:
\vspace{-0.5em}
\begin{small}
\begin{align*}
    \cG(\cA; \cD) &:= \ex{\substack{\mathbf{y}\sim \cD_y, \\ \hat{\mathbf{y}}\sim \cA(\mathbf{y})}}{\ell(f({\mathbf{X}},\mathbf{y}),f({\mathbf{X}},\hat{\mathbf{y}}))} \\
&~= \frac{1}{n^2\sigma_{\bf{y}}^2}\textstyle \sum_{k=1}^{m} \left(  \ex{\mathbf{y}, \hat{\mathbf{y}}}{\textstyle\sum_{i=1}^{n}\Tilde{x}_{ik}^2\left(y_i-\hat{y}_i\right)^{2}  + \textstyle \sum_{i=1}^n \textstyle \sum_{\substack{j=1, \\j \neq i}}^n \Tilde{x}_{ik}\Tilde{x}_{jk} (y_i - \hat{y}_i)(y_j - \hat{y}_j)}\right)\\
\end{align*}
\end{small}

\vspace{-2em}
By appropriately selecting variables, minimizing the aforementioned loss function over the parameters $\Mij{y}{\hat{y}}=\pr{\cA(y)=\hat{y}}$ indexed by $(y,\hat{y}) \in \cY \times \hat{\cY}$, which represent the phenotype randomizer, can be framed as a quadratic programming (QP) problem:

\begin{equation}\label{eqn:GWAS_qp}
\begin{aligned}
\text{minimize}_{{\bf{w}}} \quad & \mathbf{b}^T\mathbf{w} + \mathbf{w}^T\mathbf{A}\mathbf{w}, \\
\textrm{subject to}  \quad & w_u \succeq 0, \quad &\forall u \in \cY, \\
  \quad &\mathbf{1}^T w_u = 1, \quad   &\forall u \in \cY, \\
  \quad &w_u - e^{\epsilon}w_{u'} \preceq 0, \quad  &\forall u,u' \in \cY, u \neq u'.
\end{aligned}
\end{equation} 

We provide the full derivation in Supplementary Note~\hyperref[{appdx:phenoQP}]{S3}, including the construction of $\mathbf{b}$ and $\mathbf{A}$.
Note that ${\bf{w}}$ is a column vector of length $|\cY|\cdot|\hat{\cY}|$, which is formed by stacking the variables in $\Mij{y}{\hat{y}}$, with $w_u$ representing the conditional distribution given the input phenotype $u\in\cY$.
Additionally, ${\mathbf{X}}$, $\mathcal{Y}$, and $\hat{\mathcal{Y}}$ are assumed to be fixed and known. 

We obtain our $\epsilon$-DP randomizer as a solution to this optimization problem, which intuitively minimizes the expected difference between the true GWAS statistics and those calculated using the randomized phenotypes under the constraint that the randomization function satisfies the phenotypic DP constraints. 
The following states that this optimization problem has a solution (proof in Supplementary Note~\hyperref[{appdx:phenoQP}]{S3}).
\begin{proposition}\label{prop:qp_sol}
 For any prior distribution $\cD$ over $\mathbb{R}$ with finite support, and $\epsilon > 0$, there exists an $\epsilon$-DP randomizer $\cA_* \in \inf \cG(\cA;\cD)$, where the infimum is over all $\epsilon$-DP mechanisms $\cA$ described by the optimization problem in Eq.~\ref{eqn:GWAS_qp}, and $\cA_*$ outputs a set $\hat{\cY}$  with at most $|\cY|$ phenotype values. 
\end{proposition}


The optimization problem in Eq.~\ref{eqn:GWAS_qp} is a \emph{non-convex} QP, which is hard to solve in general.
Hence, we reformulate it as a difference-of-convex (DC) problem to solve it approximately using an iterative application of convex optimization methods \cite{DCA}. 
Specifically, we further rewrite the block matrix $\bf{A}$ as the difference of two block matrices $\bf{C}$ and $\bf{E}$, where each is a positive semidefinite (PSD) matrix (Supplementary Note~\hyperref[{appdx:phenoQP}]{S3}).
Let $g({\bf{w}}) = {\bf{w}}^T{\bf{C}}{\bf{w}} + {\bf{b}}^T{\bf{w}}$ and $h({\bf{w}})={\bf{w}}^T\mathbf{E}{\bf{w}} $, then the optimization can be written in the DC form as
\begin{equation}\label{eqn:qp_dca}
\begin{aligned}
\text{minimize}_{{\bf{w}}} \quad & g({\bf{w}}) - h({\bf{w}}),  \\
\textrm{subject to}  \quad & w_u \geq 0, \quad &\forall u \in \cY, \\
  \quad &1^T w_u = 1, \quad   &\forall u \in \cY, \\
  \quad &w_u - e^{\epsilon}w_u' \leq 0, \quad  &\forall u,u' \in \cY, u \neq u'.
\end{aligned}
\end{equation}
The functions $g$ and $h$ are convex and differentiable. Let $\cC$ denote the above constraint set.
Replacing $h$ with its first-order Taylor expansion at ${\bf{w}}^{(k)}$, we can iteratively solve the following convex subproblem~\cite{DCA}:
\begin{align*}
    {\bf{w}}^{(k+1)} \in \mathsf{argmin}_{{\bf{w}} \in \cC} \left(g({\bf{w}}) - \left(h({\bf{w}}^{(k)})+ \langle \nabla h({\bf{w}}^{(k)}),{\bf{w}}-{\bf{w}}^{(k)}\rangle \right) \right).
\end{align*}

\vspace{-0.5em}
To manage the cross-terms in $\cG(\cA; \cD)$, we assume that the phenotypes are jointly distributed as $\mathbf{y} \sim \cN(\mu\cdot \mathbf{1}_{n\times 1},\sigma_a^2 \mathbf{K})$. Here, $\mu$ represents the privately estimated mean derived from the original $\mathbf{y}$ vector. The term $\mathbf{K} = \Tilde{\mathbf{X}}\Tilde{\mathbf{X}}^T / m$ is commonly referred to as the genetic relatedness matrix (GRM) with an associated variance parameter $\sigma_a^2$.
Moreover, this distributional assumption incorporates genetic similarities among all pairs of individuals, thereby aligning more closely with the GWAS model.
The above QP formulation leverages the genotype matrix $\mathbf{X}$, particularly through the GRM, to model correlations between individuals' phenotypes $y_i$ and $y_j$.  However, the prior over $\mathbf{y}$ does not fully utilize the genotype information. Specifically, the algorithm assumes a multivariate Gaussian prior over $\mathbf{y}$ with identical marginal distributions for each $y_i$, independent of $\mathbf{X}$. As a result, when the phenotype values $y_i$ are (almost) independent, the QP formulation effectively reduces to GOPHER-LP:

\begin{proposition}\label{prop:lptoqp}
Let $\mathcal{D}_x$ be any distribution over the genotypes and let $\mathcal{D}_y$ be a distribution over phenotypes $\mathcal{Y}$ such that the phenotype values $y_i \in \mathbf{y}$ are independent. Then, for any $\epsilon > 0$, the randomizer $\mathcal{A}_{\mathrm{LP}}$ obtained by GOPHER-LP (Algorithm~\ref{alg:gopher-lp}) is an optimal solution to the quadratic minimization problem $\mathsf{inf}_{\cA}\ \cG(\cA; \cD_x,\cD_y)$.
\end{proposition}
This observation suggests that GOPHER-LP can effectively maximize GWAS accuracy when individuals in the cohort are largely independent. However, in many real-world studies, substantial relatedness often exists within the cohort. 

We introduce GOPHER-MultiQP (Algorithm~\ref{alg:gopher-multiqp}) to address this more general setting.
MultiQP combines the personalization strategy of MultiLP with the QP formulation to explicitly account for sample relatedness. Similar to MultiLP, we estimate the fixed-effect coefficients $\boldsymbol{\beta}$ and $\sigma^2_y$, resulting in $\mathbb{P}[y_i | \mathbf{x}_i] = \mathcal{N}(\boldsymbol{\beta}^\top \mathbf{x}_i, \sigma_y^2)$ as the personalized marginal priors for each individual. To model phenotype correlations within a group of individuals, we define the joint prior distribution as 
$\mathbb{P}[\mathbf{y}| \mathbf{X}] = \cN( \mathbf{X}\boldsymbol{\beta},\sigma^2 \mathbf{K})$. We leverage these correlated personalized priors and apply the QP formulation to groups of individuals with similar phenotype priors, yielding adaptive randomizers that both maximizes statistical accuracy in the presence of relatedness and incorporates genotype-informed beliefs over the phenotypes.

\vspace{-0.5em}
\section{Results}
\vspace{-0.5em}
\label{sec:exp}

We evaluated the GOPHER algorithms on both real and simulated phenotypes together with the genomic data from the UK Biobank to assess performance both in real study settings with noisy observations and under tunable heritability. 
For real phenotypes, we obtained three key blood lipid traits: HDL cholesterol (HDL-C), LDL cholesterol (LDL-C), and triglycerides (TG). 
These biomarkers are widely studied in cardiovascular research and exhibit modest single nucleotide polymorphism (SNP)-based heritability (Pan-UKBB~\cite{panukbb} estimates: $\hat{h}_g^2 = 0.23$ for HDL-C, $0.09$ for LDL-C, $0.18$ for TG), representing a challenging setting for GWAS with DP.
Simulated traits were generated from a standard linear additive model~\cite{regenie} across three heritability levels ($h_g^2=$ 0.2, 0.5, or 0.8).
We binned the phenotype values into a specified number of bins to apply our mechanisms; 
all experiments use $b=80$ bins within each phenotype range. 
While we sampled 100,000 subjects of white British ancestry and 500,000 SNPs for analysis, our mechanisms support releasing the statistics of all variants with no additional privacy cost, as they can be generated from the same perturbed phenotypes.

In all experiments, we used PLINK~\cite{plink}, a widely used tool for GWAS, to compute the association statistics for both the original and privatized phenotypes for comparison. 
For GOPHER-MultiLP and GOPHER-MultiQP, we estimate the phenotype prediction model using PRS-CS~\cite{PRScs} on a subset of the dataset (35k samples for real, 20k for simulated). 
Additionally, in all experiments we reserved a small portion of the privacy budget ($\epsilon_1 = 0.1$) for privately estimating the phenotype mean and variance parameters needed for constructing the phenotype priors. 
Unless otherwise specified, the privacy budget provided to the main optimization procedure remains the same across all experiments ($\epsilon-\epsilon_1$).  
GOPHER-LP and GOPHER-MultiLP experiments were conducted on an Amazon EC2 m5d.12xlarge instance with 48 Intel Xeon Platinum 8175M or 8259CL vCPUs and 192~GB RAM. GOPHER-MultiQP was run on an Amazon EC2 g4dn.16xlarge instance, equipped with 64 vCPUs from an Intel Xeon Platinum 8259CL multi-processor system and 256~GB RAM.

We illustrate the overall workflow of the GOPHER mechanisms in Supplementary Figure \ref{fig:workflow} and provide further details on data preprocessing and phenotype simulation in Supplementary Note~\hyperref[{appdx:addlDet}]{S4}. In addition, we include a parameter robustness analysis in Supplementary Table~\ref{tab:param_training_sz}, which demonstrates that GOPHER's performance remains stable across a wide range of settings.

\begin{table}[t]
\small
\begin{center}
    \begin{tabular}{|c|c|c|c|c|c|c|c|c|c|c|c|c|c|}
        \hline
        \multirow{4}{*}{\textbf{Privacy budget}} & \multirow{4}{*}{\textbf{Mechanism}} & \multicolumn{12}{c|}{\textbf{GWAS Accuracy}} \\ 
        \cline{3-14}  & & \multicolumn{2}{c|}{HDL-C}& \multicolumn{2}{c|}{LDL-C } & \multicolumn{2}{c|}{Triglycerides} & \multicolumn{6}{c|}{Simulated Phenotypes}\\
        \cline{9-14} & & \multicolumn{2}{c|}{$(\hat{h}^2_g = 0.23)$} & \multicolumn{2}{c|}{$(\hat{h}^2_g = 0.09)$} & \multicolumn{2}{c|}{$(\hat{h}^2_g = 0.18)$}& \multicolumn{2}{c|}{$h^2_g = 0.2$}& \multicolumn{2}{c|}{$h^2_g = 0.5$} & \multicolumn{2}{c|}{$h^2_g = 0.8$} \\
         \cline{3-14} & & MSE & $r$ & MSE & $r$ & MSE & $r$ & MSE & $r$ & MSE & $r$ & MSE & $r$  \\
        \hline

        \multirow{4}{*}{$\epsilon = 1.0$} & Laplace & 2.29&0.06&2.10&0.07&2.20&0.07 &2.28 & 0.08 & 2.71 & 0.11 & 3.41 & 0.12\\
        & RR & 2.39&0.02&2.24&0.01&2.35&0.01 & 2.49 & 0.001 & 2.97 & 0.01 &  3.73 & 0.02 \\
        & GOPHER-LP  & 1.56&0.38&1.55&0.32&1.51&0.37 & 2.10& 0.40 & 1.90 & 0.45 & 1.85 & 0.52 \\
        & GOPHER-MultiLP &  \textbf{1.45}& \textbf{0.43}& \textbf{1.50}& \textbf{0.36}& \textbf{1.47}& \textbf{0.40} &  \textbf{1.77} & \textbf{0.72} & \textbf{1.54} & \textbf{0.74} & \textbf{1.47} & \textbf{0.79} \\
        & GOPHER-MultiQP  & \textbf{0.53}&\textbf{0.91}&\textbf{0.65}&\textbf{0.80}&\textbf{0.55}&\textbf{0.74} &  \textbf{1.70} & \textbf{0.74} & \textbf{1.51} & \textbf{0.78} & \textbf{1.21} & \textbf{0.83} \\
        \hline
          \multirow{4}{*}{$\epsilon = 3.0$} & Laplace & 1.91&0.22&1.74&0.24&1.82&0.24 & 1.83 & 0.28 &   2.24&  0.34 & 2.57 & 0.38 \\
        & RR &2.24&0.08&2.08&0.07&2.20&0.07 & 2.27 & 0.08 &2.88 & 0.11 & 3.33 & 0.14 \\
        & GOPHER-LP &0.54&0.80&0.51&0.79&0.61&0.76 &0.52 & 0.82 & 0.55 & 0.87 & 0.58 & 0.89 \\
        & GOPHER-MultiLP &  \textbf{0.52}& \textbf{0.83}& \textbf{0.49}& \textbf{0.81}& \textbf{0.58}& \textbf{0.68}  & \textbf{0.50} & \textbf{0.84} &\textbf{0.39} & \textbf{0.94} & \textbf{0.34} & \textbf{0.93} \\
        & GOPHER-MultiQP & \textbf{0.45}&\textbf{0.87}&\textbf{0.41}&\textbf{0.86}&\textbf{0.46}&\textbf{0.85} &  \textbf{0.45} & \textbf{0.88} & \textbf{0.33} & \textbf{0.96} & \textbf{0.30} & \textbf{0.96} \\
        \hline
        \multirow{4}{*}{$\epsilon = 5.0$} & Laplace  & 1.59&0.36&1.42&0.38&1.48&0.39 & 1.44& 0.44& 1.73 & 0.51 & 1.94 & 0.56 \\
        & RR & 1.68&0.32&1.50&0.34&1.71&0.28  & 1.48 & 0.43 & 1.78 & 0.50 &  1.96 & 0.56 \\
        & GOPHER-LP & 0.21&0.92&0.18&0.93&0.24&0.91 & 0.17 & 0.94 &0.16 & 0.96 & 0.17 & 0.97 \\
        & GOPHER-MultiLP &  \textbf{0.19}& \textbf{0.93}& \textbf{0.16}& \textbf{0.94}& \textbf{0.19}& \textbf{0.93}  & \textbf{0.15} & \textbf{0.95} & \textbf{0.12} & \textbf{0.97} & \textbf{0.11} & \textbf{0.98} \\
        & GOPHER-MultiQP & \textbf{0.17}&\textbf{0.94}&\textbf{0.12}&\textbf{0.95}&\textbf{0.17} &\textbf{0.93} &  \textbf{0.13} & \textbf{0.96} & \textbf{0.10} & \textbf{0.98} & \textbf{0.09} & \textbf{0.99} \\
        \hline
    \end{tabular}
\vspace{0.5em}
\caption{\textbf{Comparison of DP mechanisms for GWAS on a UK Biobank dataset with both real and simulated phenotypes.} For each mechanism, privacy budget $\epsilon \in \{1.0, 3.0, 5.0 \}$, and phenotype, we report the mean squared error (MSE) and Pearson correlation coefficient ($r$) between the released and ground-truth $t$-statistics. GOPHER-MultiLP and MultiQP consistently achieve lower error and higher correlation, particularly under stronger privacy constraints (lower $\epsilon$). Top two performances are highlighted in bold.}
\label{tab:err}
\end{center}
\vskip -0.5in
\end{table}

\begin{figure}
\begin{center}
\vspace{1em}
\includegraphics[width=.75\columnwidth]{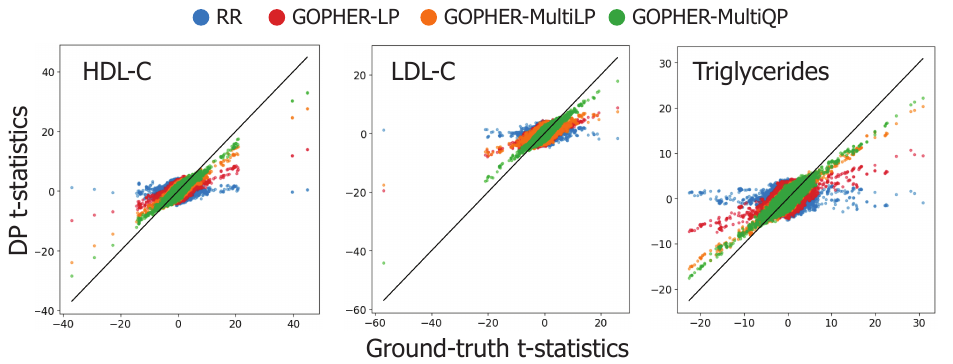}
\vspace{-0.5em}
\caption{\textbf{Comparison of differentially-private versus ground-truth association statistics.} Scatter plots compare the $t$-statistics released by each DP mechanism to the true GWAS results for HDL-C, LDL-C, and triglycerides in the UK Biobank cohort ($n=100,000$; $\epsilon = 1.0$). Closeness to the identity line shows the accuracy of each approach. }
\label{fig:eps_plot}
\end{center}
\vskip -0.25in
\end{figure}

\vspace{1em}
\noindent\textbf{GOPHER enables the accurate release of genome-wide association statistics with privacy across phenotypes of varying heritability.}
\begin{modtext}
We evaluated our approaches to phenotype randomization against widely adopted standard DP methods, including Laplace and Randomized Response (RR) mechanisms. 
Table~\ref{tab:err} presents a quantitative comparison across varying privacy budgets and heritability levels, using two accuracy metrics: mean squared error (MSE) and the Pearson correlation coefficient ($r$) between the privatized and ground-truth GWAS $t$-statistics. 
Across all three lipid traits and all tested $\epsilon$ values (i.e., privacy levels), our mechanisms consistently achieved lower MSE and higher correlation than the baseline mechanisms, with GOPHER-MultiQP achieving the best performance in all settings.
Even under the particularly challenging low-heritability and high-noise conditions of these traits (e.g., LDL-C with $\hat{h}_g^2=0.09$), our methods yield high accuracy (e.g., MSE of 0.65 for MultiQP at $\epsilon=1.0$).
Figure~\ref{fig:eps_plot} further shows the relationship between privatized versus true $t$-statistics for $\epsilon=1.0$ (see Supplementary Figure~\ref{fig:scatter_real_pheno} for other $\epsilon$ values), demonstrating that our accuracy improvements apply to variants across varying association strengths.
\end{modtext}

Our simulated phenotype experiments further highlight the benefits of personalization under controlled settings (Table~\ref{tab:err} and Supplementary Figure~\ref{fig:scatter_sim_pheno}).
Across all privacy budgets, GOPHER-MultiLP and GOPHER-MultiQP outperformed GOPHER-LP, which optimizes the mechanism without personalization.
The utility gains of MultiQP and MultiLP over LP are particularly pronounced at higher heritability ($h^2_g = 0.8$), where phenotypes are more strongly associated with genotypes, leading to more accurate personalized priors compared to the global prior.

\begin{figure}[t]
\begin{center}
\centerline{\includegraphics[width=\columnwidth]{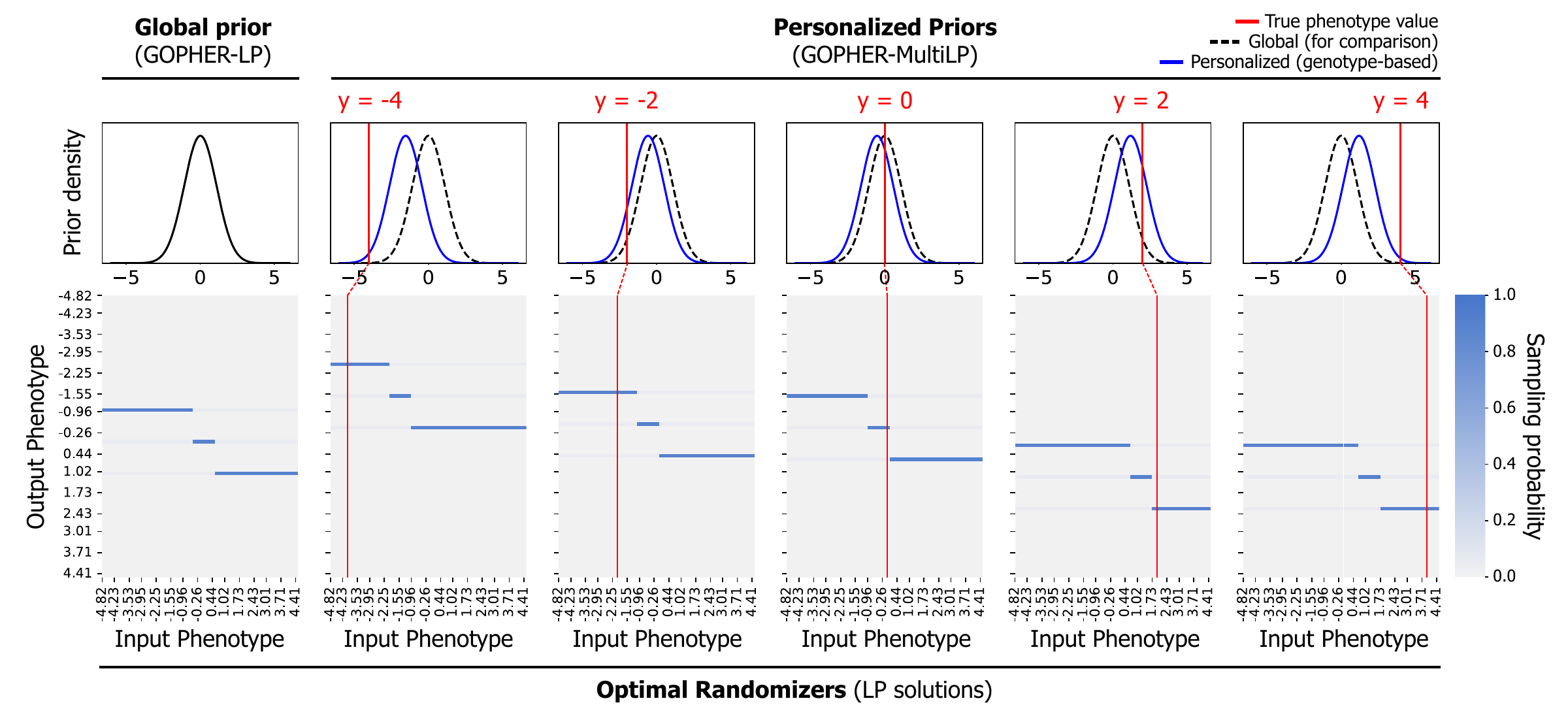}}
\vspace{-0.5em}
\caption{\textbf{The impact of personalized priors on the optimal randomizers.} Global prior (left) used by GOPHER-LP results in a single randomizer across all individuals; conditional distributions of the randomizer are visualized as a heatmap on the bottom. In contrast, GOPHER-MultiLP leverages individual-specific priors (right) predicted based on genotypes (in blue), resulting in randomizers better aligned with the expected phenotypes. Examples from five UK Biobank individuals are shown.}
\label{fig:prior_plot}
\end{center}
\vskip -0.5in
\end{figure}

Figure~\ref{fig:prior_plot} illustrates the impact of individual-specific priors by showing how the GOPHER-MultiLP mechanism randomizes phenotype values compared to GOPHER-LP. For example, with $\epsilon =3$ and $h^2_g=0.8$, LP selects three fixed bins around $-1.03$, $-0.02$, and $1.01$, mapping the sensitive phenotype values to these bins with high probability. In contrast, MultiLP assigns sensitive phenotype values to various bins based on personalized priors so that the joint distribution of the original and randomized labels is more consistent with the estimated (predicted) phenotype prior. 


To assess performance under realistic patterns of genetic relatedness, we included 32,500 pairs of third-degree or closer relatives in our GWAS cohort, as determined by the estimated kinship coefficients provided by the UK Biobank. The observation that GOPHER-MultiQP consistently outperforms GOPHER-MultiLP across all tested privacy budgets (Table~\ref{tab:err}) indicates that optimizing the statistical accuracy objective via a quadratic program provides an additional layer of improvement beyond personalized priors, enabled by more faithfully capturing relatedness within the cohort and its impact on the phenotype distributions.

\vspace{1em}
\noindent\textbf{Accuracy and runtime across varying dataset sizes.}
We conducted additional benchmark experiments to assess GOPHER's performance across varying sample sizes and privacy budgets (Supplementary Figure~\ref{fig:sam_plot}). As expected, accuracy improves with larger datasets because less noise is required to satisfy a given privacy level. Notably, even at a sample size of 25k, GOPHER-MultiLP achieves a low MSE of approximately 0.75 at $\epsilon = 2$, indicating that our mechanisms can provide accurate GWAS results while maintaining meaningful privacy protection even for moderately sized cohorts.

The wall-clock runtimes of our mechanisms (Supplementary Table~\ref{tab:runtime}) are generally practical, ranging from under 10 seconds for GOPHER-LP to roughly 10 minutes for GOPHER-MultiLP, and approximately 6.5 hours for GOPHER-MultiQP on a cohort of 100,000 individuals. The main computational bottleneck in GOPHER-MultiQP stems from constructing the QP block matrices, a step that scales quadratically with the sample size $(n^2)$. However, our optimized implementation, which leverages caching of redundant Gaussian density evaluations (Supplementary Note~\hyperref[{appdx:comp_complx}]{S5}), substantially reduces this cost, requiring only about 2 hours for 100,000 samples. Extrapolating from these results, we estimate a feasible runtime of approximately 2.5 days for a large-scale dataset of 500,000 individuals, illustrating the broad practical applicability of our approach.

\vspace{1em}
\noindent\textbf{Comparison with existing differentially private GWAS approaches.}
\begin{modtext}
We compared GOPHER-MultiLP and GOPHER-MultiQP to prior DP approaches for GWAS, which generally add noise directly to the released statistics rather than randomizing phenotypes as in GOPHER. Under the standard Laplace mechanism, each SNP is perturbed independently for DP, causing noise to accumulate linearly with the number of released SNPs and leading to poor recovery of top associations~\cite{johnson2013privacy,uhlerop2013privacy,simmons2016realizing}. PrivSTRAT~\cite{PrivSTRAT} improves on this approach by supporting population stratification correction and tighter sensitivity analysis (i.e., more precise calculation of the required amount of noise), yielding better accuracy in heterogeneous populations. However, because PrivSTRAT’s accuracy degrades rapidly as $K$ increases, it has only been evaluated for small values of $K$ (3 or 5). In contrast, GOPHER removes dependence on $K$ and provides accurate DP statistics for all variants, enabling post hoc identification of top associations without incurring additional privacy cost.

We evaluated GOPHER, PrivSTRAT, and the na\"{i}ve Laplace mechanism for releasing top-$K$ statistics using the same UK Biobank cohort with real phenotypes, applying correction for five inferred ancestry components when computing the ground-truth statistics to maintain consistency with the target setting addressed by PrivSTRAT. 
GOPHER also offers a straightforward way to account for population stratification by including ancestry covariates in the association testing step following phenotype randomization.
We measured top-$K$ accuracy, defined as the proportion of true top-$K$ SNPs correctly recovered. Across all phenotypes at $\epsilon = 3$, GOPHER-MultiLP and GOPHER-MultiQP achieve high top-$K$ accuracy over a wide range of $K$, substantially outperforming prior approaches (Figure~\ref{fig:prior_dpgwas}). For example, at $K = 100$, MultiLP and MultiQP recover 94 and 95 true top associations, respectively, compared with fewer than 10 recovered by both PrivSTRAT and Laplace baselines. This improvement is further reflected in the MSE of the released statistics, where GOPHER achieves 5--10 times lower MSE across all phenotypes (Supplementary Figure~\ref{fig:prior_dpgwas_mse}).

\begin{figure}[t]
\begin{center}
\includegraphics[width=.75\columnwidth]{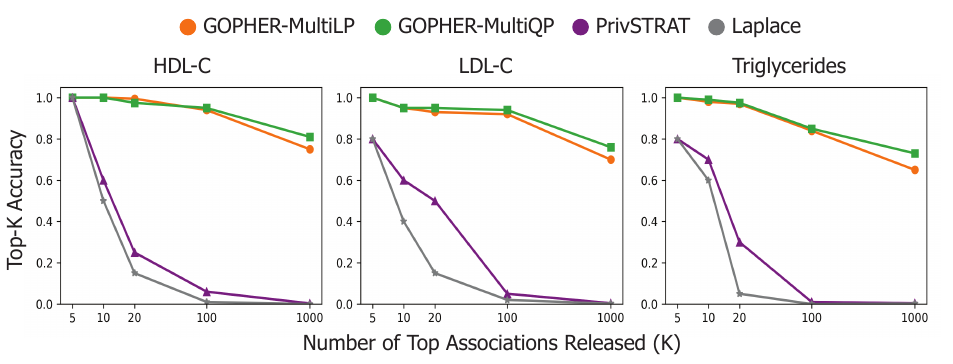}
\vspace{-0.5em}
\caption{\textbf{Accuracy of releasing top-$K$ associations under different DP mechanisms.} We compare the top-$K$ accuracy across GOPHER-MultiLP, GOPHER-MultiQP, PrivSTRAT, and Laplace mechanisms ($\epsilon = 3$) on the UK Biobank cohort for HDL-C, LDL-C, and Triglycerides. GOPHER maintains high accuracy across a wide range of $K$, while accuracy of prior methods declines sharply as $K$ increases due to cumulative noise from independent DP releases.}
\label{fig:prior_dpgwas}
\end{center}
\vskip -0.25in
\end{figure}
\end{modtext}

\vspace{-0.5em}
\section{Conclusions and Future Work}
\vspace{-0.5em}
\begin{modtext}
We presented GOPHER mechanisms for releasing full GWAS statistics with enhanced utility under phenotypic differential privacy.  By integrating optimization-based randomization and personalized priors, GOPHER enables the accurate release of full set of genome-wide statistics with privacy showcasing performance well beyond what was previously possible with existing approaches. Although we focused on quantitative traits in our study, our mechanisms naturally extend to the randomization of binary traits.

Beyond the technical contributions, our methods have important implications for responsible genomic data sharing. Privacy-preserving approaches like GOPHER can strengthen public trust by safeguarding participant data, thereby encouraging broader participation in future studies. Although often viewed as impractical in biomedical settings, our work demonstrates that, with careful design, differential privacy can be operationally feasible while providing meaningful privacy guarantees for genomic data sharing.

Regarding future directions, one limitation of our current framework is the composition of differential privacy across multiple phenotypes, which complicates privacy budget allocation when releasing GWAS results for several traits from the same cohort. Addressing this challenge will require improved modeling of covariance among related phenotypes to reduce the noise required in multi-trait settings. Another challenge is ensuring compatibility between privatized GWAS outputs and downstream tools, as perturbed statistics may violate model assumptions and introduce bias. Adapting downstream analyses to account for the structured noise introduced by GOPHER may therefore be important for applications such as meta-analysis, fine-mapping, and risk modeling. Future work may also refine our optimization objectives to better align with specific downstream analyses. Overall, our work provides a practical foundation for developing tailored DP mechanisms to support secure genomic data sharing.

\end{modtext}

\begin{credits}
\subsubsection{\ackname} 
This work is partially supported by National Institutes of Health Early Independence Award DP5 OD029574 and National Science Foundation award 2452612 (to H.C.). This research has been conducted using the UK Biobank resource under applications 93997 and 41910. Data access applications can be submitted via the UK Biobank portal (\url{https://www.ukbiobank.ac.uk}).

\subsubsection{\discintname}
The authors have no competing interests to declare that are relevant to the content of this paper.  
\end{credits}
\bibliographystyle{unsrt}
\bibliography{ref}
%





\newpage
\begin{appendix}
\setcounter{page}{1}
\renewcommand{\appendixname}{Supplementary Note}
\renewcommand{\thesection}{Supplementary Note S\arabic{section}}
\setcounter{section}{0}
\renewcommand{\thefigure}{S\arabic{figure}}
\renewcommand{\figurename}{Supplementary Figure}
\setcounter{figure}{0}
\renewcommand{\tablename}{Supplementary Table} \renewcommand{\thetable}{S\arabic{table}}
\setcounter{table}{0}

\clearpage
\section{Additional Related Work}\label{appdx:priv-strat}
\begin{modtext}
A related line of work, exemplified by I-GWAS~\cite{igwas}, follows an “exact-but-incomplete” paradigm that assumes privacy risks are negligible if a predefined safety condition holds, based on the combinatorial space for reconstructing genotypes. However, this relies on strong and often unrealistic assumptions about allele frequencies and data distributions, provides no individual-level privacy guarantees under targeted attacks, and limits output to a small subset of variants, excluding rare variants and those in linkage disequilibrium, thereby reducing utility for fine-mapping and personalized medicine. In contrast, our phenotype randomization approach offers rigorous differential privacy guarantees, decouples noise from the number of variants, and scales efficiently. It also naturally extends to I-GWAS settings with overlapping cohorts, allowing randomized phenotypes to be reused across multiple GWAS analyses without incurring additional privacy loss, due to the post-processing property of DP. 

Other recent label-DP methods target different applications and are not directly applicable to GWAS release, but we briefly discuss them for completeness. LabelDP-Pro~\cite{labeldp-pro} improves DP-SGD utility by denoising gradients using non-private features, conceptually similar to our use of PRS models but specialized to deep learning. Badanidiyuru et al.\cite{badanidiyuru2023} introduce unbiasedness constraints in randomized response over bins; while compatible with our approach, such constraints may reduce utility since we directly optimize expected error in the final GWAS statistics. Esfandiari et al.\cite{esfandiari2022label} use clustering-based label-DP, but in GWAS, clustering tends to reflect ancestry, information that has limited predictive value for most phenotypes, making it less effective.
\end{modtext}

\clearpage

\section{Review of Randomized Response on Bins}\label{appdx:rr-on-bins}
We briefly summarize a related prior work of Ghazi et al.~\cite{ghazi2023regression}, which introduces an optimal label DP mechanism for a given objective loss function. 
Formally, for a loss function $\ell : \cX \times \cY \rightarrow \mathbb{R}$, the noisy label loss of a randomizer $\cA$ with respect to a marginal prior $\cD_y$ is given by $\ex{y \sim\cD_y,\hat{y}\sim\cA(y)}{\ell(y,\hat{y})}$, where $\hat{y}$ the noisy label generated by $\cA$ on input $y$.
 The optimal mechanism minimizing this loss can be found by solving the following linear program (LP) (cf. Equation~2, \cite{ghazi2023regression}).
\begin{equation} \label{eqn:rr-on-bins}
    \begin{aligned} \nonumber
    \text{minimize}_{M} \quad & \sum_{u \in \cY}\pr{u} \left(\sum_{v \in \hat{\cY}} \Mij{u}{v}\cdot\ell(u,v)\right), \\  
    \textrm{subject to} \quad & \forall  u  \in \cY,v \in \hat{\cY}: \quad\quad  \Mij{u}{v} \geq 0,  \\
      & \forall u \in \cY: \quad   \sum_{v \in \hat{\cY}} \Mij{u}{v} = 1,   \\ 
      & \forall v \in \hat{\cY}, \forall u,u' \in \cY, u \neq u', : \quad \Mij{u'}{v} \leq e^{\epsilon_2} \Mij{u}{v}. 
    \end{aligned}
    \end{equation}

\begin{definition}[Randomized Response on Bins]\label{def:rr-on-bins}
Let $\epsilon> 0$, for map $\phi: \cY \rightarrow \hat{\cY}$,  $\textsf{RR-on-Bins}^\phi_\epsilon$ is defined as the mechanism that on input $y$ samples $\hat{y} \sim \hat{Y}$, where the random variable $\hat{Y}$ is distributed as:
\begin{align*}
 \pr{\hat{Y} =\hat{y}} =
    \begin{cases}
       \frac{e^\epsilon}{e^\epsilon+|\hat{\cY}|-1} , & \text{for } \hat{y} = \phi(y), \\
        \frac{1}{e^\epsilon+|\hat{\cY}|-1},              & \text{otherwise.}
    \end{cases}
\end{align*}
\end{definition}
The following theorem summarizes the main result of Ghazi et al.~\cite{ghazi2023regression}:
\begin{theorem}
For any loss function $\ell: \cY \times \hat{\cY} \rightarrow \mathbb{R}$, all finitely supported distributions $\cP$ over $\cY$, there is an output set $\hat{\cY}$ and a non-decreasing map $\phi: \cY \rightarrow \hat{\cY}$ such that $\cL(\textsf{RR-on-Bins}^\phi_\epsilon; \cP) = \inf_{\cA} \cL(\cA; \cP)$; where the infimum is over all $\epsilon$-DP mechanism $\cA$ described by the LP given in Equation~\ref{eqn:rr-on-bins}.
\end{theorem}

This result implies that, given a global prior on label values and a desired privacy level $\epsilon$, one can construct an optimal label-DP mechanism that minimizes expected loss for common regression metrics such as Poisson log loss, mean squared error, and mean absolute error. Notably, the optimal mechanism takes the form of \emph{randomized response on bins} (Definition~\ref{def:rr-on-bins}), and the authors provide an efficient algorithm (Algorithm~2) to determine the optimal binning strategy.



\clearpage
\section{Detailed Derviation and Analysis of MultiQP}\label{appdx:phenoQP}
In this section, we give the omitted full derivation and proofs from Section~\ref{subsec:multiqp}. Recall that GOPHER-MultiQP is designed to directly optimize the accuracy of the association statistics—specifically, the Pearson correlation coefficient. Let $f(\Tilde{\mathbf{X}},\mathbf{y}) = \frac{\Tilde{\mathbf{X}}^T(\mathbf{y-\mu_{\mathbf{y}}})}{n \sigma_{\mathbf{y}}} $ denote the vector of Pearson correlation coefficients between the standardized genotype matrix $\tilde{\mathbf{X}} \in \boldsymbol{\cX}^n \subset \bR^{n \times m}$ and the phenotype vector $\mathbf{y} \in \cY^n \subset \mathbb{R}^n$. Let $\cD$ be an unknown distribution over $\boldsymbol{\cX} \times \cY$, i.e., $(\mathbf{x}_i, y_i) \sim \mathcal{D}$, and let $\cD_\mathbf{y}$ denote the marginal distribution over $\mathbf{y}$. Let $\hat{\mathbf{y}}$ denote the privatized phenotype vector obtained by applying an $\epsilon$-DP mechanism $\mathcal{A}$ to $\mathbf{y}$.

Our objective is to minimize the expected error in the target statistics due to privatization (randomization). We can expand this objective as follows:

\small{
\begin{align*}
    &~ \ex{\substack{\mathbf{y}\sim \cD_\mathbf{y}, \\ \hat{\mathbf{y}}\sim \cA(\mathbf{y})}}{\ell(f({\mathbf{X}},\mathbf{y}),f({\mathbf{X}},\hat{\mathbf{y}}))} \\
    = &~  \ex{\substack{\mathbf{y}\sim \cD_\mathbf{y}, \\ \hat{\mathbf{y}}\sim \cA(\mathbf{y})}}{\frac{1}{n^2\sigma_\mathbf{y}^2}\|\Tilde{\mathbf{X}}^T(\mathbf{y}-\hat{\mathbf{y}})\|_{F}^{2}} \\
    = &~ \ex{\substack{\mathbf{y}\sim \cD_\mathbf{y}, \\ \hat{\mathbf{y}}\sim \cA(\mathbf{y})}}{\frac{1}{n^2\sigma_\mathbf{y}^2}\left(\tr\left(\Tilde{\mathbf{X}}^T(\mathbf{y}-\hat{\mathbf{y}})(\mathbf{y}-\hat{\mathbf{y}})^T\Tilde{\mathbf{X}}\right)\right)}\\
    = &~ \frac{1}{n^2\sigma_\mathbf{y}^2}\sum_{k=1}^{m} \ex{\mathbf{y}, \hat{\mathbf{y}}}{\mathbf{\Tilde{x}}_{*,k}^T(\mathbf{y}-\hat{\mathbf{y}})(\mathbf{y}-\hat{\mathbf{y}})^T\mathbf{\Tilde{x}}_{*,k}}\\
  = &~  \frac{1}{n^2\sigma_{\bf{y}}^2}\sum_{k=1}^{m}\ex{\mathbf{y}, \hat{\mathbf{y}}}{ \left( \sum_{i=1}^{n} \Tilde{x}_{ik}(y_i-\hat{y}_i) \right)^{2}}   \\
= &~  \frac{1}{n^2\sigma_{\bf{y}}^2}\sum_{k=1}^{m} \left(  \ex{\mathbf{y}, \hat{\mathbf{y}}}{\sum_{i=1}^{n}\Tilde{x}_{ik}^2\left(y_i-\hat{y}_i\right)^{2}  + \sum_{i=1}^n \sum_{j=1,j \neq i}^n \Tilde{x}_{ik}\Tilde{x}_{jk} (y_i - \hat{y}_i)(y_j - \hat{y}_j)}\right)\\
= &~ \frac{1}{n^2\sigma_{\bf{y}}^2}\sum_{k=1}^{m} \left(\ex{\mathbf{y}}{\ex{\hat{\mathbf{y}}|\mathbf{y}}{ \sum_{i=1}^{n} \Tilde{x}_{ik}^2\left(y_i-\hat{y}_i\right)^{2} + \sum_{i=1}^n \sum_{j=1,j \neq i}^n \Tilde{x}_{ik}\Tilde{x}_{jk} (y_i - \hat{y}_i)(y_j - \hat{y}_j)}}  \right).
\end{align*}}

\normalsize
We let $\Mij{y}{\hat{y}}=\pr{\cA(y)=\hat{y}}$ denote the conditional output probabilities of the $\epsilon$-DP mechanism $\mathcal{A}$, and  $\pr{y_i}$ denote the marginal prior distribution for each individual. Then, the first (diagonal) term in the nested expectation simplifies to:
\begin{align*}
\ex{\mathbf{y}}{\ex{\hat{\mathbf{y}}|\mathbf{y}}{ \sum_{i=1}^{n} \Tilde{x}_{ik}^2 \left(y_i-\hat{y}_i\right)^{2}}} &= \sum_{i=1}^{n} \Tilde{x}_{ik}^2 \sum_{u\in \cY} \pr{y_i =u} \sum_{v\in \hat{\cY}} \Mij{u}{v}(u - v)^2. 
\end{align*}
\noindent Let $\mathbb{P}_{\mathbf{y}}(y_i,y_j)$ denote the joint distribution over $(y_i, y_j)$. The second (off-diagonal) expectation term becomes:
\begin{align*}
&\ex{\mathbf{y}}{\ex{\hat{\mathbf{y}}|\mathbf{y}}{\sum_{i=1}^n \sum_{j=1,j \neq i}^n \Tilde{x}_{ik}\Tilde{x}_{jk}(y_i - \hat{y}_i)(y_j - \hat{y}_j)}}  = \ex{\mathbf{y}}{\sum_{i=1}^n \sum_{j=1,j \neq i}^n \Tilde{x}_{ik}\Tilde{x}_{jk} \ex{\hat{y}_i|y_i}{y_i - \hat{y}_i} \cdot \ex{\hat{y}_j|y_j}{y_j - \hat{y}_j} }\\
&=\ \ex{\mathbf{y}}{\sum_{i=1}^n \sum_{j=1,j \neq i}^n \Tilde{x}_{ik}\Tilde{x}_{jk} \left(\sum_{v \in \hat{\cY}} \Mij{y_i}{v} (y_i - v)\right) \cdot \left(\sum_{v' \in \hat{\cY}} \Mij{y_j}{v'} (y_j - v')\right)}\\
&=\ \sum_{i=1}^n \sum_{j=1,j \neq i}^n \Tilde{x}_{ik}\Tilde{x}_{jk} \sum_{u \in \cY}\sum_{u' \in \cY} \mathbb{P}_{\mathbf{y}}(y_i=u,y_j=u')\left(\sum_{v \in \hat{\cY}} \Mij{u}{v} (u - v)\right) \cdot \left(\sum_{v' \in \hat{\cY}} \Mij{u'}{v'} (u' - v')\right). 
\end{align*}

\noindent To simplify the above terms, we define the following matrices for all ${\bf{x}}_{i} \in \boldsymbol{\cX}, u\in \cY$ and  $v \in \hat{\cY}$: 
$$ \quad  P_{{\bf{y}}}(i,u) = \pr{y_{i}=u}, \  M_{\hat{\mathbf{y}}|\mathbf{y}}(u,v) = \Mij{u}{v},\ \text{ and } \ D_{\mathbf{y}-\hat{\mathbf{y}}}(u,v)= u - v.$$
\noindent Using these definitions, we can rewrite the expected loss as:
\small
\begin{align*}
    &\mathbb{E}_{\mathbf{y},\hat{\mathbf{y}}}\left[\ell\left(f(\mathbf{X},\mathbf{y}),f(\mathbf{X},\hat{\mathbf{y}})\right)\right]  =\frac{1}{n^2\sigma_y^2}\sum_{k=1}^{m} \biggl( \sum_{i=1}^{n} \Tilde{x}_{ik}^2 \sum_{u \in \cY} P_\mathbf{y}(i,u) (D_{\mathbf{y}-\hat{\mathbf{y}}}(u,\cdot) \circ D_{\mathbf{y}-\hat{\mathbf{y}}}(u,\cdot))M_{\hat{\mathbf{y}}|\mathbf{y}}(u,\cdot)^T  \\
    & \quad\quad\quad +   \sum_{i=1}^n \sum_{j=1,j \neq i}^n \Tilde{x}_{ik}\Tilde{x}_{jk} \sum_{u \in \cY} \sum_{u' \in \cY}  \mathbb{P}_{\mathbf{y}}(y_i=u,y_j=u') D_{\mathbf{y}-\hat{\mathbf{y}}}(u,\cdot)M_{\hat{\mathbf{y}}|\mathbf{y}}(u,\cdot)^T M_{\hat{\mathbf{y}}|\mathbf{y}}(u',\cdot) D_{\mathbf{y}-\hat{\mathbf{y}}}(u',\cdot)^T  \biggr),
\end{align*}
\normalsize
where $\circ$ denotes element-wise multiplication. Further, we can vectorize the expression as:
\small
\begin{align*}
    \ex{\mathbf{y},\hat{\mathbf{y}}}{\ell(f(\mathbf{X},\mathbf{y}),f(\mathbf{X},\hat{\mathbf{y}}))}   = &\frac{1}{n^2\sigma_y^2} \biggl(\sum_{k=1}^{m} \sum_{u \in \cY} (\mathbf{\Tilde{x}}_{*,k} \circ \mathbf{\Tilde{x}}_{*,k})^T P_\mathbf{y}(\cdot,u) (D_{\mathbf{y}-\hat{\mathbf{y}}}(u,\cdot) \circ D_{\mathbf{y}-\hat{\mathbf{y}}}(u,\cdot))M_{\hat{\mathbf{y}}|\mathbf{y}}(u,\cdot)^T  \\
    &  +  \sum_{u \in \cY} \sum_{u' \in \cY} M_{\hat{\mathbf{y}}|\mathbf{y}}(u',\cdot) D_{\mathbf{y}-\hat{\mathbf{y}}}(u',\cdot)^T C(u,u') D_{\mathbf{y}-\hat{\mathbf{y}}}(u,\cdot)M_{\hat{\mathbf{y}}|\mathbf{y}}(u,\cdot)^T   \biggr),
    \end{align*}
    \normalsize
    where
    \begin{align*}
        C(u,u') =  &\sum_{k=1}^{m} \sum_{i=1}^n \sum_{j=1,j \neq i}^n \Tilde{x}_{ik}\Tilde{x}_{jk} \mathbb{P}_{\mathbf{y}}(y_i=u,y_j=u').
\end{align*}

We now express the loss function in a compact quadratic form. Let  $w_u \in \bR^{|\hat{\cY}|}$ denote the column vector corresponding to the input value $u$ in the (transposed) matrix $M_{\hat{\mathbf{y}}|\mathbf{y}}^T$, i.e., $w_u[v] =\pr{\cA(u)=v}, ~\forall v \in \hat{\cY} $. Let ${\bf{w}} = \textsf{vec}(M_{\hat{\mathbf{y}}|\mathbf{y}}^T) \in \bR^{|\cY|.|\hat{\cY}|}$ denote the stacked vector formed by vertically concatenating all such $w_u$ for $u \in \cY$. 

We define the following two variables for our formulation of the optimization problem.
\paragraph{Linear term.} Define the vector $\mathbf{b}$ of length $|\cY|\cdot|\hat{\cY}| $, as the concatenation of vectors $b_u$, one for each $u \in \cY$, where:
 $$b_u = \sum_{k=1}^{m} (\Tilde{\bf{x}}_{*,k} \circ \Tilde{\bf{x}}_{*,k})^T P_{{\bf{y}}}(\cdot,u) (D_{\mathbf{y}-\hat{\mathbf{y}}}(u,\cdot) \circ D_{\mathbf{y}-\hat{\mathbf{y}}}(u,\cdot)).$$
\paragraph{Quadratic term. } Similarly, let $\mathbf{A}$ be a $(|\cY|\cdot|\hat{\cY}|) \times (|\cY|\cdot|\hat{\cY}|)$ block matrix comprised of a total of $|\cY|^2$ individual matrices. Each component matrix within this arrangement is denoted as $A_{uu'}$ and is defined as follows:
\begin{align*}
A_{uu'} &= C(u,u')\cdot \left(D_{\mathbf{y}-\hat{\mathbf{y}}}(u',\cdot)^T D_{\mathbf{y}-\hat{\mathbf{y}}}(u,\cdot)\right), \quad \forall u,u' \in \cY.
\end{align*}

\paragraph{Final optimization problem.} Now we can express minimizing the expected squared error in the vector of correlation statistics as the following quadratic programming (QP) problem:
\begin{equation*}
\begin{aligned}
\min_{{\bf{w}}} \quad & \mathbf{b}^T\mathbf{w} + \mathbf{w}^T\mathbf{A}\mathbf{w} \\
\textrm{subject to}  \quad & w_u \succeq 0, \quad &\forall u \in \cY, \\
  \quad &\mathbf{1}^T w_u = 1, \quad   &\forall u \in \cY, \\
  \quad &w_u - e^{\epsilon}w_{u'} \preceq 0, \quad  &\forall u,u' \in \cY, u \neq u'.
\end{aligned}
\end{equation*}

The above optimization problem is non-convex because the matrix $\mathbf{A}$ in the quadratic term is not guaranteed to be positive semidefinite (PSD). This is due to the presence of cross-terms  in $C(u,u')$, which encode covariance across individuals, and depending on the structure of the data, they can lead to negative eigenvalues in $\mathbf{A}$.
Since a non-convex quadratic program (QP) is generally hard to solve, we reformulate the problem as a difference-of-convex (DC) program and solve it approximately using the DC algorithm (DCA) \cite{DCA}, which iteratively linearizes the concave part and solves a sequence of convex subproblems. 

We decompose the matrix 
$\mathbf{A}$ into the difference of two PSD matrices: $\mathbf{A} = \mathbf{C} -  \mathbf{E}$, where ${\bf{C}}= {\bf{A}}+ \rho\mathbb{I}_{(|\cY|\cdot|\hat{\cY}|)}$ and ${\bf{E}} = \rho\mathbb{I}_{(|\cY|\cdot|\hat{\cY}|)}$, with $\rho>0$ chosen to ensure that $\mathbf{C}$ is PSD ($\mathbf{E}$ is PSD by design) and $\mathbb{I}_d$ denoting the $d\times d$ identity matrix.  In practice, we set $\rho = \max(0,-\lambda_{\min}({\bf{A}}))$, where $\lambda_{\min}({\bf{A}})$ denotes the smallest eigenvalue of $\mathbf{A}$. This allows us to rewrite the objective as: 
\begin{equation}
\begin{aligned}\nonumber
\min_{{\bf{w}}} \quad & g({\bf{w}}) - h({\bf{w}})  \\
\textrm{subject to}  \quad & w_u \succeq 0, \quad &\forall u \in \cY, \\
  \quad &\mathbf{1}^T w_u = 1, \quad   &\forall u \in \cY, \\
  \quad &w_u - e^{\epsilon}w_{u'} \preceq 0 \quad  &\forall u,u' \in \cY, u \neq u'.
\end{aligned}
\end{equation}
where, $g({\bf{w}}) = {\bf{w}}^T{\bf{C}}{\bf{w}} + {\bf{b}}^T{\bf{w}}$ and $h({\bf{w}})={\bf{w}}^T\mathbf{E}{\bf{w}} $. 


Both $g$ and $h$ are convex and differentiable. Let $\cC$ denote the constraint set given in the problem above. 
Now replacing $h$ with its first-order Taylor expansion at ${\bf{w}}^{(k)}$, the DCA algorithm iteratively solves the following convex subproblem:
$$ {\bf{w}}^{(k+1)} \in \mathsf{argmin}_{{\bf{w}} \in \cC} \left(g({\bf{w}}) - \left(h({\bf{w}}^{(k)})+ \langle \nabla h({\bf{w}}^{(k)}),{\bf{w}}-{\bf{w}}^{(k)}\rangle \right) \right).$$
Simplifying, we get:
\begin{small}
\begin{align*}
     {\bf{w}}^{(k+1)} &\in \mathsf{argmin}_{{\bf{w}} \in \cC} \left( {\bf{w}}^T{\bf{C}}{\bf{w}} + {\bf{b}}^T{\bf{w}} - \left(({\bf{w}}^{(k)})^T\mathbf{E}{\bf{w}}^{(k)}+ \langle 2\mathbf{E}{\bf{w}}^{(k)},{\bf{w}}-{\bf{w}}^{(k)}\rangle \right) \right)\\
  &= \mathsf{argmin}_{{\bf{w}} \in \cC} \left( {\bf{w}}^T{\bf{C}}{\bf{w}} + {\bf{b}}^T{\bf{w}} -  \langle 2\mathbf{E}{\bf{w}}^{(k)},{\bf{w}}\rangle - ({\bf{w}}^{(k)})^T\mathbf{E}{\bf{w}}^{(k)}+ \langle 2\mathbf{E}{\bf{w}}^{(k)},{\bf{w}}^{(k)}\rangle  \right)\\   
   &= \mathsf{argmin}_{{\bf{w}} \in \cC} \left( {\bf{w}}^T{\bf{C}}{\bf{w}} + \left({{\bf{b}} - 2\mathbf{E}{\bf{w}}^{(k)}}\right)^T{\bf{w}}  + ({\bf{w}}^{(k)})^T\mathbf{E}{\bf{w}}^{(k)}  \right).
\end{align*}
\end{small}

\noindent Thus, each DCA iteration reduces to solving a convex QP over the constraint set $\cC$. We initialize $\mathbf{w}^{(0)}$ using the solution to the linear part of the objective, i.e., by solving the convex optimization problem that minimizes only the linear term $\mathbf{b}^T\mathbf{w} $ subject to the differential privacy and probability simplex constraints. This provides a feasible and informative starting point that aligns with minimizing per-coordinate label noise, and we then apply DCA iterations until convergence.

We now formalize the guarantees of this procedure by proving the key propositions stated earlier. First, we establish in Proposition~\ref{prop:qp_sol} that the GOPHER-MultiQP algorithm converges to a local minimizer of the quadratic objective over the space of $\epsilon$-differentially private mechanisms, under standard conditions on the objective and constraint set. Following this, we prove Proposition~\ref{prop:lptoqp}, which shows that when phenotype values are independent across individuals, the  quadratic problem decomposes into a set of simpler linear programs, allowing the GOPHER-MultiLP mechanism to be optimal in application settings where the samples are expected to be independent.
\paragraph{Proof of Proposition~\ref{prop:qp_sol}}
Since $\cD$ has finite support and $\cY$ is finite, the conditional probability matrix $\Mij{y}{\hat{y}}$ can be represented as a point in a finite-dimensional space $\mathbb{R}^{|\cY| \times |\hat{\cY}|}$. The differential privacy constraints and probability normalization conditions define a closed and bounded feasible set, which is therefore compact.

The objective in Eq.~\ref{eqn:GWAS_qp} is a continuous function of $\Mij{y}{\hat{y}}$, and due to compactness of the feasible set, an optimal solution exists. Additionally, since the objective is convex in $\Mij{y}{\hat{y}}$ and the feasible region is convex (with linear inequality and equality constraints), the optimal solution can be expressed as a convex combination of at most $|\cY| \times |\hat{\cY}|$ extreme points by Carathéodory’s theorem \cite{barvinok2002course}. Furthermore, by results in  (e.g.,~\cite{ghosh2012,ghazi2023regression}), there exists an optimal mechanism \( \mathcal{A}_* \) whose output support \( \hat{\mathcal{Y}} \) has size at most \( |\mathcal{Y}| \). Therefore, there exists an optimal mechanism $\cA_*$ whose output support $\hat{\cY}$ has size at most $|\cY|$.
To solve the optimization problem in Eq.~\ref{eqn:GWAS_qp}, which is a non-convex quadratic program we use the Difference-of-Convex algorithm (DCA).
DCA reformulates the objective as a difference of two convex functions, $f({\bf{w}}) = g({\bf{w}}) - h({\bf{w}})$, and iteratively solves convex subproblems of the form:
\begin{align*}
    {\bf{w}}^{(k+1)} \in \mathsf{argmin}_{{\bf{w}} \in \cC} \left(g({\bf{w}}) - \left(h({\bf{w}}^{(k)})+ \langle \nabla h({\bf{w}}^{(k)}),{\bf{w}}-{\bf{w}}^{(k)}\rangle \right) \right),
\end{align*}
where $\cC$ is the feasible set defined by the DP constraints. 

As shown in ~\cite{tao1997convex}, under mild conditions such as continuity and differentiability of $g$ and $h$, the sequence ${\mathbf{w}^{(k)}}$ generated by DCA converges to a critical point of the objective. Although DCA does not guarantee convergence to a global optimum, it is known to perform well in practice for structured non-convex problems and often achieves high-quality local optima efficiently.

\paragraph{Proof of Proposition~\ref{prop:lptoqp}}
The key assumption is that the phenotype values $\{y_i\}_{i=1}^n$ are independent under $\mathcal{D}_\mathbf{y}$, i.e., $\pr{\mathbf{y}} = \prod_i \pr{y_i}$. Therefore, the joint distribution over $(y_i, y_j)$ factorizes:
\begin{small}
\begin{align*}
&\ex{\mathbf{y}}{\ex{\hat{\mathbf{y}}|\mathbf{y}}{\sum_{i=1}^n \sum_{j=1,j \neq i}^n \Tilde{x}_{ik}\Tilde{x}_{jk}(y_i - \hat{y}_i)(y_j - \hat{y}_j)}}  = \sum_{i=1}^n \sum_{j=1,j \neq i}^n \Tilde{x}_{ik}\Tilde{x}_{jk} \ex{y_i}{\ex{\hat{y}_i|y_i}{y_i - \hat{y}_i}}  \ex{y_i}{\ex{\hat{y}_j|y_j}{y_j - \hat{y}_j} }.
\end{align*}
\end{small}

This implies that all the second-order (cross) terms in the loss can be written as outer products of expectations over single individuals. Hence, the quadratic terms become separable in $i$, and the global minimization problem decomposes into a sum of independent local minimization problems, one for each $y_i$.
Therefore, the GOPHER-LP mechanism, which solves the local LP for each individual (using global $\epsilon$-DP constraints), minimizes each term in the decomposed sum independently and hence minimizes the overall loss.

\begin{modtext}
\paragraph{GOPHER-MultiQP.} Unlike MultiLP, we do not apply GOPHER-QP independently to each individual, as this approach would be computationally expensive and and impractical for large-scale datasets. Instead, we reduce the number of QPs by grouping individuals whose marginal prior distributions are similar. Accordingly, we partition the data into $k$ disjoint clusters $\{(\mathbf{X}_2^{(1)}, \mathbf{y}_2^{(1)}),\ldots, (\mathbf{X}_2^{(k)}, \mathbf{y}_2^{(k)})\}$, where each cluster $(\mathbf{X}_2^{(j)}, \mathbf{y}_2^{(j)})$ contains a subset of data samples from $(\mathbf{X}, \mathbf{y})$. To construct these clusters, we apply the $k$-means clustering algorithm to the mean vector $\mathbf{X\beta}$, obtained from the estimated personalized prior distribution. This ensures that labels with similar feature distributions are grouped together. 

For each cluster, we construct a dedicated quadratic program (QP) following the same formulation as in the original GOPHER-QP setup. These cluster-specific QPs are then combined into a unified QP with a block-structured form that captures dependencies across clusters while maintaining computational tractability.

Let ${\bf{w}}^{[j]} $ denote the optimization variable corresponding to cluster  $(\mathbf{X}_2^{(j)}, \mathbf{y}_2^{(j)})$, and let $\mathbf{A}^{[j]}$ denote the matrix $\mathbf{A}$ constructed using only the samples from cluster $j$. Similarly, let $\mathbf{A}^{[k,j]}$ denote the matrix constructed using the samples from both clusters $k$ and $j$. Define $\mathbf{b}^{[j]}$ as the corresponding subvector of $\mathbf{b}$. We then define the full set of optimization variables and constraints in the overall formulation as follows:
\begin{equation*}
\begin{aligned}
    \mathbf{\Tilde{w}} =\begin{bmatrix}
        \mathbf{w}^{[1]}\\
        \mathbf{w}^{[2]}\\
        \vdots\\
        \mathbf{w}^{[k]}\\
    \end{bmatrix} ,
    \quad
    \mathbf{\Tilde{A}} = \begin{bmatrix}
        \mathbf{A}^{[1]} & \mathbf{A}^{[1,2]} & \ldots & \mathbf{A}^{[1,k]} \\
        \mathbf{A}^{[1,2]^T} & \mathbf{A}^{[2]} & \ldots & \mathbf{A}^{[2,k]} \\
        \mathbf\vdots & \vdots & \ldots & \vdots \\
        \mathbf{A}^{[1,k]^T} & \mathbf{A}^{[2,k]^T} & \ldots & \mathbf{A}^{[k]} \\
    \end{bmatrix},
    \quad
    \mathbf{\Tilde{b}} =\begin{bmatrix}
        \mathbf{b}^{[1]}\\
        \mathbf{b}^{[2]}\\
        \vdots\\
        \mathbf{b}^{[k]}\\
    \end{bmatrix}.
\end{aligned}
\end{equation*}
The resulting clustered QP (MultiQP) is then formulated as:
\begin{equation}\label{eqn:multiqp}
\begin{aligned}
\min_{\mathbf{\Tilde{w}}} \quad & \mathbf{\Tilde{b}}^T\mathbf{\Tilde{w}} + \mathbf{\Tilde{w}}\mathbf{\Tilde{A}}\mathbf{\Tilde{w}} \\
\textrm{subject to}  \quad & w_u^{[j]} \succeq 0, \quad &\forall u \in \cY, \forall j \in [k], \\
  \quad &\mathbf{1}^T w_u^{[j]} = 1, \quad   &\forall u \in \cY,\forall j \in [k], \\
  \quad &w_u^{[j]} - e^{\epsilon}w_{u'}^{[j]} \preceq 0, \quad  &\forall u,u' \in \cY, u \neq u',\forall j \in [k].
\end{aligned}
\end{equation}
This formulation unifies the individual cluster-specific QPs into a single comprehensive optimization problem, while retaining the structural properties and constraint relationships inherent to each cluster.
\end{modtext}

\newpage
\section{Additional Experimental Details}\label{appdx:addlDet}
Here we outline the data processing pipeline used to evaluate the GOPHER mechanisms under differential privacy. We describe the preprocessing steps applied to both genotype and phenotype data, the quality control procedures, and the simulation methodology used to generate synthetic traits from real genetic datasets.
To adhere to standard quality control criteria~\cite{PRScs}, we excluded samples with a mismatch between genetically inferred sex and self-reported sex, high genotype missingness, extreme heterozygosity, sex chromosome aneuploidy, and those excluded from kinship inference and autosomal phasing prior to sampling. Additionally, only single nucleotide polymorphisms (SNPs), the most common type of genetic variant, from the HapMap3 database were considered, after removing ambiguous SNPs and markers with minor allele frequency (MAF) $<1\%$, missing rate $>1\%$, and imputation quality INFO score $<0.8$, as well as those that significantly deviated from Hardy-Weinberg equilibrium with $p\text{-value} > 10^{-10}$. Missing values were imputed using the Haplotype Reference Consortium dataset. This process resulted in a genotype dataset comprising 96 million variants and 100,000 individuals. We sampled 500,000 SNPs for analysis; however, the mechanism is directly applicable to the full set of SNPs without requiring additional privacy budget.
\paragraph{Phenotype simulation.} We sampled phenotypes using a standard approach found in the literature~\cite{regenie} based on real genotypes. Specifically, we selected $P=100$ variants across all chromosomes to be considered causal variants. For each causal SNP, the effect size $\beta_j$ was sampled from a normal distribution with mean $0$ and variance $h^2_g$ divided by the number of causal variants. Here, $h^2_g$ denotes the heritability explained by genome-wide genetic markers, commonly referred to as narrow-sense heritability, which we fixed to $0.5$ or $0.8$. Subsequently, we employed a linear model to generate the phenotype as 
$y_i = \sum_{j=1}^P x_{i,j}\beta_j + e_i,$
where $x_{i,j}$ represents the standardized genotype of individual $i$ at SNP $j$, and $e_i$ is a normally distributed noise variable with a mean of $0$ and variance of $1-h^2_g$, representing environmental effects.
Handling phenotypic outliers is a critical step in GWAS analyses to minimize false positives and ensure data reliability. In our experiments, we addressed extreme phenotypic values by clipping the range of the privatized phenotypes to the top and bottom $5\%$ boundaries, computed using the differentially private mean and standard deviation estimates. 

For GOPHER-MultiQP, the number of clusters $k$ was fixed at the maximum feasible value of $7$, constrained by the CVXOPT optimization library’s limits on parameter dimensionality. Empirical results indicate that this configuration yields robust performance, although exploring larger $k$ values with alternative solvers may further enhance model flexibility.

\clearpage
\section{Computational Complexity of MultiQP}\label{appdx:comp_complx}



\begin{modtext}
 
For GOPHER-MultiQP, constructing the matrices in the QP problem  has  computational complexity $O(k^2 n^2 b^2)$, where $n$ is the number of individuals, $k$ is the number of clusters, and $b$ is the number of discretization bins. Solving the QP itself has an asymptotic complexity of $O((b^2 k)^3)$, but in practice this step is inexpensive, taking less than four hours even for large cohorts when using the recommended setting of $k=7$. The main computational bottleneck is therefore the construction the QP block matrices, which is highly parallelizable and amenable to engineering optimizations. 

The dominant cost arises from evaluating joint probability distributions over discretized grids of $\mathbf{y}$ values, scaling as $O(b^2)$ per sample pair. Within each cluster, the computation scales as $O(n_i (u + n_j) b^2)$, where $n_i$ and $n_j$ denote the cluster dimensions and $u$ is the number of unique mean and variance configurations. Using the grouping approach to estimate the prior, many phenotypes share the same mean–variance configuration. By caching Gaussian probability evaluations for these repeated configurations, redundant computations are avoided, substantially reducing wall time while the asymptotic complexity per cluster remais $O(b^2 n^2)$ .  Using these optimizations, the runtime for GOPHER-MultiQP for 100,000 individuals becomes quite practical (around 6-7 hours; see Supplementary Table~\ref{tab:runtime}), and we estimate a feasible total runtime of 2--3 days for a cohort of 500{,}000 individuals.

Since the main computational bottleneck is the construction of the QP block matrices rather than solving the QP itself, additional speedups could be achieved through approximation. In particular, because each QP instance involves relatively small matrices (e.g., $b \times b$) aggregated across all sample pairs ($n^2$), sampling-based strategies could provide accurate estimates for very large cohorts. We consider this a promising direction for future work.

\end{modtext}

\clearpage

 
\begin{table}[p]
\begin{center}
\textbf{(A) Training sample size sensitivity} \\[0.3em]
    \begin{tabular}{|c|c|c|c|c|c|c|c|}
        \hline
        \multirow{3}{*}{Privacy budget} & \multirow{3}{*}{Phenotype} & \multicolumn{6}{c|}{Training Sample Size} \\ 
        \cline{3-8} & & \multicolumn{2}{c|}{$t=20000$} &\multicolumn{2}{c|}{$t=35000$ } &\multicolumn{2}{c|}{$t=50000$} \\
        \cline{3-8} & & MSE & $r$ & MSE & $r$ &  MSE & $r$ \\
        \hline
        \multirow{2}{*}{$\epsilon = 1.0$} & Simulated & 0.63 &0.92 & 0.76 & 0.89& 0.91 & 0.83  \\
        \cline{2-8}& HDL-C  &0.84  &0.67 &0.81  &0.69 & 0.90 & 0.63  \\
        \hline
        \multirow{2}{*}{$\epsilon = 3.0$} & Simulated & 0.49& 0.95 & 0.56 & 0.94&0.61&0.93 \\
        \cline{2-8}& HDL-C &0.28 &0.90 & {0.29} &0.90 & 0.30 & 0.88\\
        \hline
        \multirow{2}{*}{$\epsilon = 5.0$} & Simulated& {0.31} &{0.98}  & 0.33&0.98  &0.36  & 0.97 \\
        \cline{2-8}& HDL-C & 0.19 & 0.93  &{0.17} & {0.94} & 0.19 & 0.93   \\
        \hline
    \end{tabular}
\vspace{1.5em}

\textbf{(B) Bin count sensitivity} \\[0.3em]
\begin{tabular}{|c|c|c|c|c|c|c|c|c|c|}
        \hline
        \multirow{3}{*}{Privacy budget} & \multirow{3}{*}{Mechanism} & \multicolumn{8}{c|}{Phenotype bin counts ($b$)} \\ 
        \cline{3-10} & & \multicolumn{2}{c|}{$50$} &\multicolumn{2}{c|}{$80$ (default)} &\multicolumn{2}{c|}{$100$} &\multicolumn{2}{c|}{$200$} \\
         \cline{3-10} & & MSE & $T$ & MSE & $T$ &  MSE & $T$ & MSE & $T$ \\
        \hline
        \multirow{2}{*}{$\epsilon = 1.0$} & GOPHER-LP &  1.55& 10.69& 1.56 & 10.83& 1.56 & 10.88&  1.56& 12.62 \\
        & GOPHER-MultiLP & 0.84 & 180& 0.81 & 540&0.82  &1290 &0.83  &5400 \\
        \hline
        \multirow{2}{*}{$\epsilon = 3.0$} & GOPHER-LP & 0.55 & 11.78 &0.55 &10.77 &0.55 & 11.06 & 0.55 & 12.66 \\
        & GOPHER-MultiLP & 0.31 & 180 & 0.30 & 420 &0.32 & 780& 0.31 & 4260 \\
        \hline
        \multirow{2}{*}{$\epsilon = 5.0$} & GOPHER-LP   &  0.21& 10.74&0.20  &10.96 & 0.22 & 11.01 & 0.22 & 12.45\\
        & GOPHER-MultiLP  & 0.18 &180 &0.17  & 480&0.17  & 1980 &0.17  & 5400\\
        \hline
    \end{tabular}

\vspace{1.5em}
\caption{\textbf{Robustness of GOPHER to choice of hyperparameters.} (A) 
We assess the sensitivity of GOPHER-MultiLP to the number of samples used for prior estimation (with a bin count of 80).  Performance remains consistent across training splits, with only minimal differences in accuracy as the number of samples allocated to prior estimation varies. Accuracy decreases slightly at $50{,}000$, likely due to the corresponding reduction in samples available for association testing. Allocating approximately $20{,}000$–$35{,}000$ individuals provides an effective balance between estimation precision and statistical power. (B) We evaluate the sensitivity of GOPHER-LP and GOPHER-MultiLP to the phenotype bin count (with training size of $35{,}000$, HDL\text{-}C). Performance is highly stable across bin counts, with MSE varying only marginally as the discretization level changes, and the default choice of $80$ bins offers an effective trade-off between runtime and phenotype resolution.  Because GOPHER-MultiLP and GOPHER-MultiQP share the same structural components, these patterns provide practical guidance for parameter choices in GOPHER-MultiQP as well. Across both analyses, performance is consistent over a broad range of settings, demonstrating the robustness and generalizability of the selected hyperparameters. All hyperparameters were fixed based on preliminary analyses on an independent simulated phenotype, and the privacy budget applies only to mechanism optimization and statistical release.}

\label{tab:param_training_sz}
\end{center}
\vskip -0.5in
\end{table}

\clearpage
\begin{table}[p]
\vskip -0.5in
\begin{center}
    \begin{tabular}{|c|c|c|}
        \hline
        \multirow{2}{*}{Mechanism} & \multirow{2}{*}{Sample size}  & {Runtime} \\
        & &   (secs)  \\
        \hline
         GOPHER-LP & 100000  & 7.8 \\
          \hline
         \multirow{4}{*} {GOPHER-MultiLP}& 10000  & 68.3  \\
         & 25000  & 162.7  \\
         & 50000 & 483.5  \\
         & 100000  & 632.8  \\
          \hline
          \multirow{4}{*} {GOPHER-MultiQP}& 10000  &   6132.0 (391.2) \\
           & 25000 &  8971.1 (883.9) \\
            & 50000  &   10881.6 (2{,}301.7) \\
         & 100000 &  23148.3 (7{,}368.3) \\
        \hline
    \end{tabular}
    

\vspace{1.5em}
\caption{\textbf{Practical runtimes of GOPHER mechanisms.} 
We report the total wall-clock runtime (in seconds) for GOPHER randomizers across varying dataset sizes. 
For GOPHER-MultiQP, the numbers in parentheses represent the runtime of the dominant computational step corresponding to the construction of the QP block matrices, which requires evaluating joint probability distributions over discretized grids. 
Solving the QP itself is computationally inexpensive, yielding a total runtime of under 7 hours for 100{,}000 individuals and an extrapolated runtime of 2--3 days for 500{,}000 individuals.
GOPHER-LP remains highly efficient due to a single global optimization, and GOPHER-MultiLP scales efficiently with increasing data size.}

\label{tab:runtime}
\end{center}
\end{table}





\clearpage

\begin{figure}[p]
\begin{center}
\centerline{\includegraphics[width=\columnwidth]{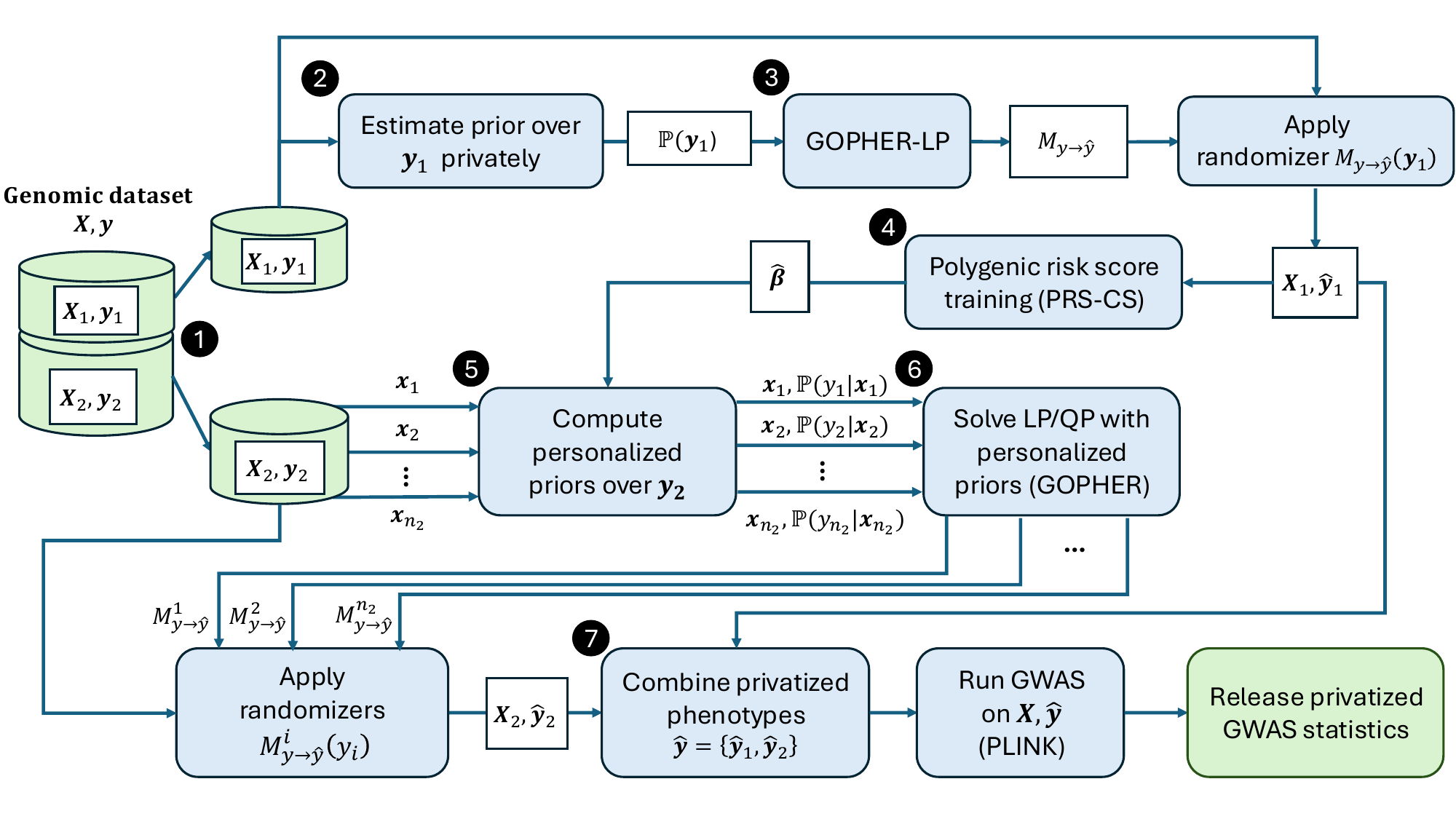}}
\caption{ \textbf{GOPHER workflow details.} This figure provides a detailed illustration of the end-to-end workflow used by GOPHER-MultiLP and GOPHER-MultiQP. We illustrate the key steps of our approach to phenotype randomization for privacy-preserving GWAS. Inputs include a genotype matrix $\mathbf{X}$, a phenotype vector $\mathbf{y}$, and a privacy budget $\epsilon=\epsilon_1+\epsilon_2$. (\textbf{1}) The dataset is randomly split into two disjoint subsets: $(\mathbf{X}_1, \mathbf{y}_1)$ for prior estimation, and $(\mathbf{X}_2, \mathbf{y}_2)$ for adaptive randomization. (\textbf{2}) We estimate a prior distribution over $\mathbf{y}_1$ using a standard private histogram method. (\textbf{3}) We apply GOPHER-LP to $(\mathbf{X}_1, \mathbf{y}_1)$ using privacy budget $\epsilon_1$ to generate privatized phenotypes $\hat{\mathbf{y}}_1$. (\textbf{4}) We train a phenotype prediction (PRS) model $P_\theta(y|\mathbf{x})$ using $(\mathbf{X}_1, \hat{\mathbf{y}}_1)$. (\textbf{5}) 
We use the trained model to compute a personalized prior distribution $P_i = P_\theta(y_i|\mathbf{x}_i)$ for each individual $i$ in $\mathbf{X}_2$.
(\textbf{6}) For each individual (or group) in the second subset, we solve the corresponding LP (MultiLP) or QP (MultiQP) using the personalized priors to obtain adaptive randomizers.  (\textbf{7}) For each $y_i$ in $\mathbf{y}_2$, we sample a privatized phenotype $\hat{y}_i \in \hat{\cY}$ from the associated randomizer. The privatized phenotypes from both subsets, $\hat{\mathbf{y}}_1$ and $\hat{\mathbf{y}}_2$, are then concatenated and used with the full genotype matrix $\mathbf{X}$ as input to standard GWAS software (e.g., PLINK) to output full GWAS results with $\epsilon$-phenotypic DP. The figure serves as a visual summary of Algorithms~\ref{alg:gopher-multilp} and \ref{alg:gopher-multiqp}.}
\label{fig:workflow}
\end{center}
\end{figure}

\clearpage
\begin{figure}[p]
\begin{center}
\vspace{-2em}
\includegraphics[width=.75\columnwidth]{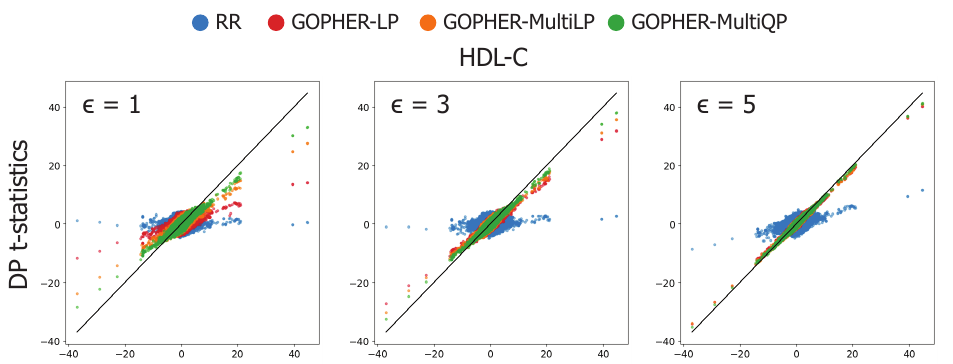}
\includegraphics[width=.75\columnwidth]{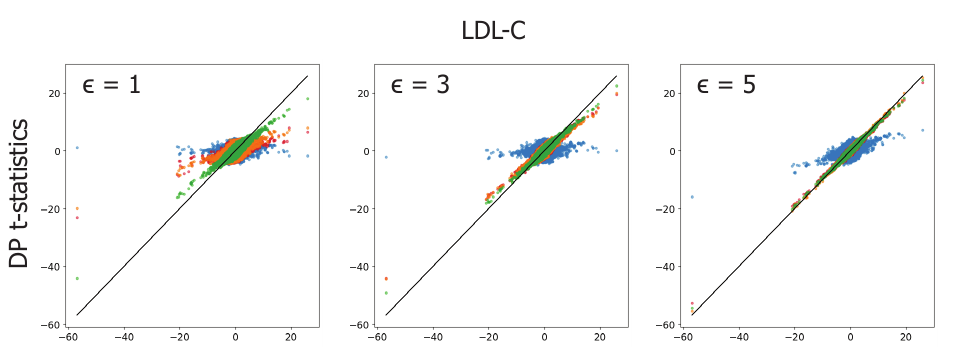}
\includegraphics[width=.75\columnwidth]{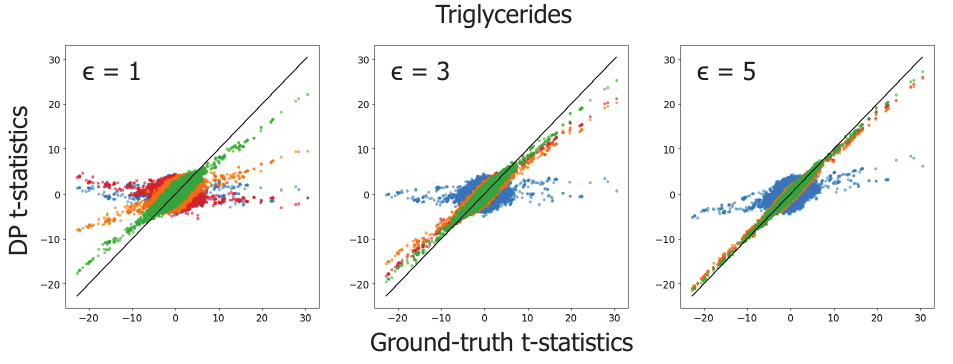}
\vspace{-0.5em}
\caption{\textbf{Differentially-private versus true statistics on the UK Biobank dataset with real phenotypes.} 
Scatter plots compare released $t$-statistics from DP mechanisms to the true GWAS results for real phenotypes (HDL-C, LDL-C, and Triglycerides) across privacy budgets $\epsilon \in \{1.0, 3.0, 5.0\}$. Proximity of points to the identity line indicates the accuracy of the differentially-private statistics.}
\label{fig:scatter_real_pheno}
\end{center}
\vskip -0.5in
\vspace{-2em}
\end{figure}

\clearpage
\begin{figure}[p]
\begin{center}
\vspace{-2em}
\includegraphics[width=.75\columnwidth]{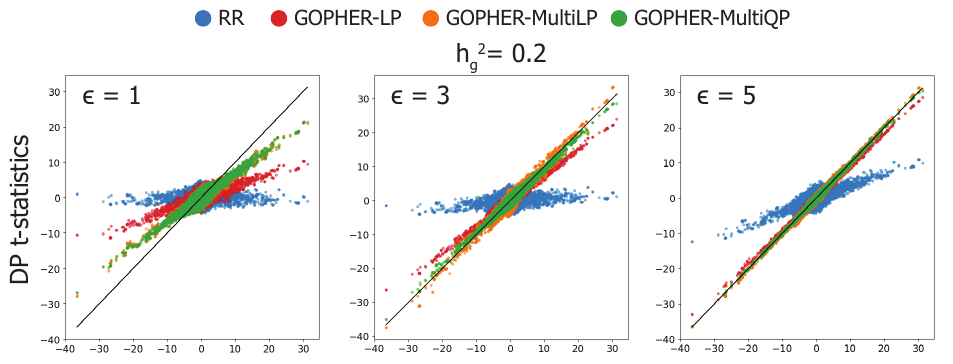}
\includegraphics[width=.75\columnwidth]{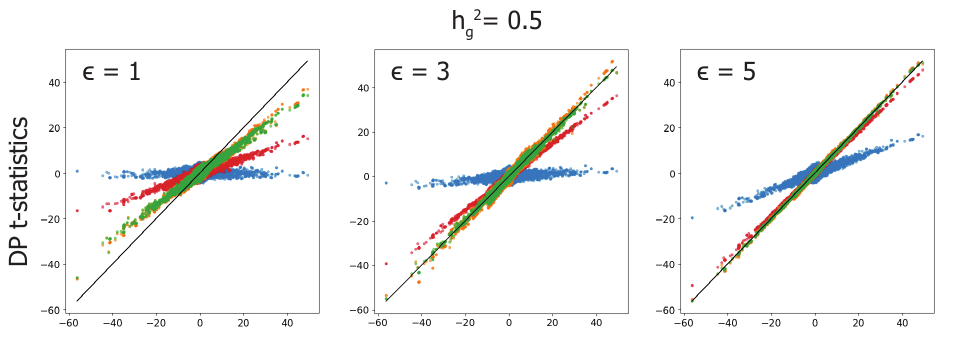}
\includegraphics[width=.75\columnwidth]{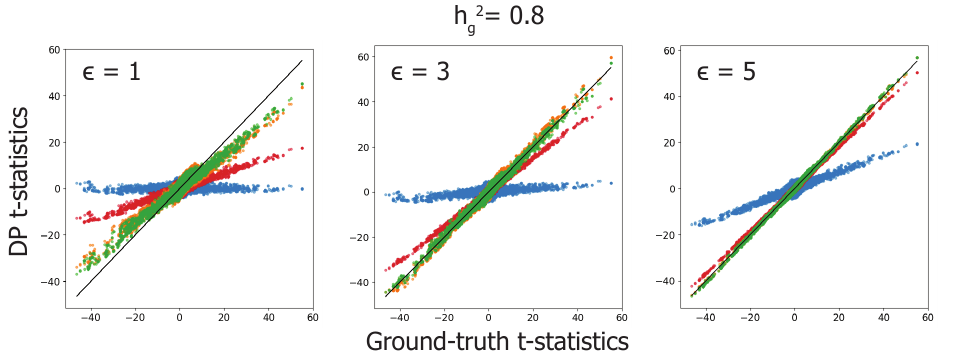}
\vspace{-0.5em}
\caption{\textbf{Differentially-private versus true statistics on the UK Biobank dataset with simulated phenotypes.} 
Scatter plots compare released $t$-statistics from DP mechanisms to the true GWAS results for simulated phenotypes across three heritability levels ($h_g^2=$ 0.2, 0.5, or 0.8) and across privacy budgets $\epsilon \in \{1.0, 3.0, 5.0\}$. Proximity of points to the identity line indicates the accuracy of the differentially-private statistics.}
\label{fig:scatter_sim_pheno}
\end{center}
\vskip -0.5in
\vspace{-2em}
\end{figure}


\clearpage

\begin{figure}[p]
\begin{center}
\includegraphics[width=.75\columnwidth]{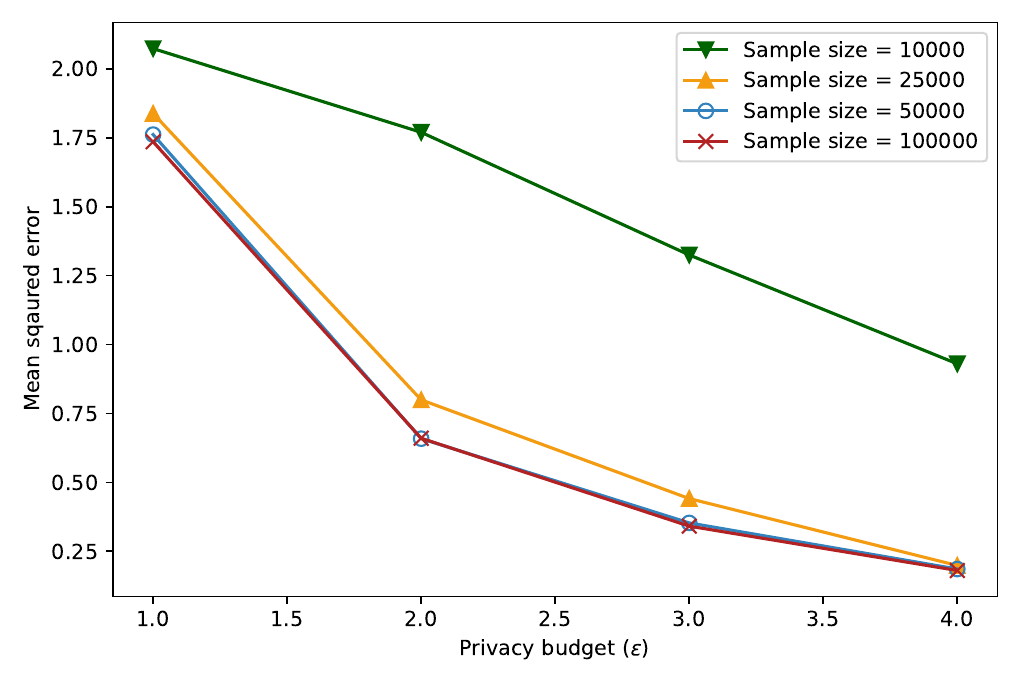}
\vspace{-0.5em}
\caption{ \textbf{Privacy-utility trade-off and scaling behavior under varying sample sizes.} We report the mean squared error (MSE) of GWAS $t$-statistics produced by GOPHER-MultiLP mechanism, across a range of privacy budgets ($\epsilon$) and sample sizes for a simulated phenotype with heritability $h^2_g=0.8$. As $\epsilon$ increases, MSE decreases for all sample sizes, with larger samples consistently yielding lower errors under the same privacy level. These results suggest that GOPHER can yield accurate privatized statistics under moderate privacy constraints, though performance degrades for smaller cohorts due to the need to allocate a subset of individuals for prior estimation. 
}
\label{fig:sam_plot}
\end{center}
\vskip -0.25in
\end{figure}

\clearpage
\begin{figure}[p]
\begin{center}
\includegraphics[width=.75\columnwidth]{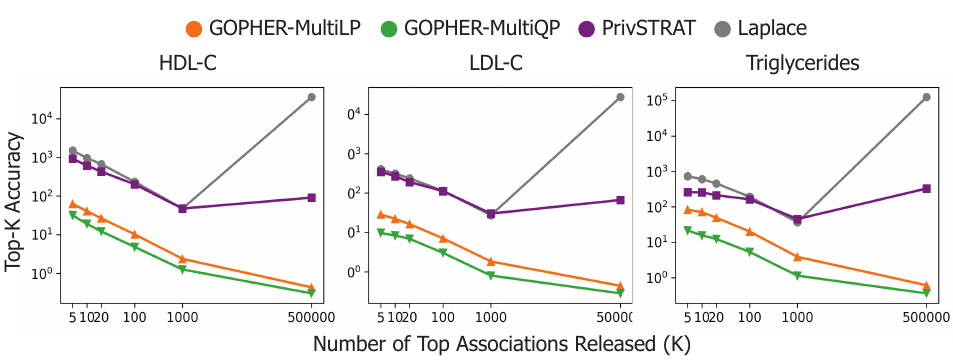}
\vspace{-0.5em}
\caption{\textbf{Mean squared error (MSE) of $t$-statistics for top-$K$ release of association statistics with differential privacy.} The x-axis shows the number of top association statistics ($K$), and the y-axis (in log-scale) shows the MSE computed over the released statistics with $\epsilon = 3$. GOPHER-MultiLP and GOPHER-MultiQP maintain low error across a wide range of $K$, while PrivSTRAT and Laplace mechanisms result in significantly larger errors across all values of $K$. 
}
\label{fig:prior_dpgwas_mse}
\end{center}
\vskip -0.25in
\end{figure}

\end{appendix}

\end{document}